\documentstyle[12pt]{article}
\input psfig.sty
\def\nn{\noindent}

\def\Re{{\cal R \mskip-4mu \lower.1ex \hbox{\it e}\,}}
\def\Im{{\cal I \mskip-5mu \lower.1ex \hbox{\it m}\,}}
\def\ie{{\it i.e.}}
\def\eg{{\it e.g.}}

\def\etal{{\it et al.}}
\def\ibid{{\it ibid}.}
\def\sub#1{_{\lower.25ex\hbox{$\scriptstyle#1$}}}
\def\tev{\,{\ifmmode\mathrm {TeV}\else TeV\fi}}
\def\gev{\,{\ifmmode\mathrm {GeV}\else GeV\fi}}
\def\mev{\,{\ifmmode\mathrm {MeV}\else MeV\fi}}
\def\to{\rightarrow}
\def\slash{\not\!}
\def\subw{_{\rm w}}
\def\mh{\ifmmode m\sbl H \else $m\sbl H$\fi}
\def\mch{\ifmmode m_{H^\pm} \else $m_{H^\pm}$\fi}
\def\mt{\ifmmode m_t\else $m_t$\fi}
\def\mc{\ifmmode m_c\else $m_c$\fi}
\def\mz{\ifmmode M_Z\else $M_Z$\fi}
\def\mw{\ifmmode M_W\else $M_W$\fi}
\def\mws{\ifmmode M_W^2 \else $M_W^2$\fi}
\def\mhs{\ifmmode m_H^2 \else $m_H^2$\fi}   
\def\mzs{\ifmmode M_Z^2 \else $M_Z^2$\fi}
\def\mts{\ifmmode m_t^2 \else $m_t^2$\fi}
\def\mcs{\ifmmode m_c^2 \else $m_c^2$\fi}
\def\mchs{\ifmmode m_{H^\pm}^2 \else $m_{H^\pm}^2$\fi}
\def\ztwo{\ifmmode Z_2\else $Z_2$\fi}
\def\zone{\ifmmode Z_1\else $Z_1$\fi}
\def\mtwo{\ifmmode M_2\else $M_2$\fi}
\def\mone{\ifmmode M_1\else $M_1$\fi}
\def\tb{\ifmmode \tan\beta \else $\tan\beta$\fi}
\def\xw{\ifmmode x\subw\else $x\subw$\fi}
\def\ch{\ifmmode H^\pm \else $H^\pm$\fi}
\def\lum{\ifmmode {\cal L}\else ${\cal L}$\fi}
\def\inpb{\,{\ifmmode {\mathrm {pb}}^{-1}\else ${\mathrm {pb}}^{-1}$\fi}}
\def\infb{\,{\ifmmode {\mathrm {fb}}^{-1}\else ${\mathrm {fb}}^{-1}$\fi}}
\def\epem{\ifmmode e^+e^-\else $e^+e^-$\fi}
\def\ppb{\ifmmode \bar pp\else $\bar pp$\fi}
\def\bsg{\ifmmode B\to X_s\gamma\else $B\to X_s\gamma$\fi}
\def\bsll{\ifmmode B\to X_s\ell^+\ell^-\else $B\to X_s\ell^+\ell^-$\fi}
\def\bstt{\ifmmode B\to X_s\tau^+\tau^-\else $B\to X_s\tau^+\tau^-$\fi}
\def\lamt{\ifmmode \tilde\lambda\else $\tilde\lambda$\fi}
\def\shat{\ifmmode \hat s\else $\hat s$\fi}
\def\that{\ifmmode \hat t\else $\hat t$\fi}
\def\uhat{\ifmmode \hat u\else $\hat u$\fi}

\newskip\zatskip \zatskip=0pt plus0pt minus0pt
\def\matth{\mathsurround=0pt}
\def\lsim{\mathrel{\mathpalette\atversim<}}
\def\gsim{\mathrel{\mathpalette\atversim>}}
\def\atversim#1#2{\lower0.7ex\vbox{\baselineskip\zatskip\lineskip\zatskip
  \lineskiplimit 0pt\ialign{$\matth#1\hfil##\hfil$\crcr#2\crcr\sim\crcr}}}

\renewcommand{\thefootnote}{\fnsymbol{footnote}}

\begin{document} \begin{titlepage} 
\rightline{\vbox{\halign{&#\hfil\cr
&SLAC-PUB-7430\cr
&March 1997\cr}}}
\begin{center}

{\Large\bf
Much Ado About Leptoquarks: A Comprehensive Analysis}
\footnote{Work supported by the Department of 
Energy, Contract DE-AC03-76SF00515}
\medskip

\normalsize 
{\large JoAnne L. Hewett and Thomas G. Rizzo } \\
\vskip .3cm
Stanford Linear Accelerator Center \\
Stanford CA 94309, USA\\
\vskip .3cm

\end{center}

\begin{abstract} 

We examine the phenomenological implications of a $\sim 200$ GeV leptoquark in 
light of the recent excess of events at HERA.  Given the relative predictions of
events rates in $e^+p$ versus $e^-p$, we demonstrate that classes of leptoquarks
may be excluded, including those contained in $E_6$ GUT models.  It is shown 
that future studies with polarized beams at HERA 
could reveal the chirality of the leptoquark fermionic coupling and that given 
sufficient luminosity in each $e^\pm_{L,R}$ channel the leptoquark quantum 
numbers could be determined.  The implications of $200-220$ GeV leptoquarks at 
the
Tevatron are examined.  While present Tevatron data most likely excludes vector
leptoquarks and leptogluons in this mass region, it does allow for scalar 
leptoquarks.  We find that while leptoquarks have little influence on Drell-Yan
production, further studies at the Main Injector may be possible in the single 
production channel provided the Yukawa couplings are sufficiently large.  
We investigate precision electroweak measurements as well 
as the process $e^+e^-\to q\bar q$ at LEP II and find they provide no further 
restrictions on these leptoquark models.  We then ascertain that cross section 
and polarization asymmetry measurements at the NLC provide the only direct 
mechanism to determine the leptoquark's electroweak quantum numbers.  The 
single production of leptoquarks in $\gamma e$ collisions by both the 
backscattered laser and Weisacker-Williams techniques at the NLC is also 
discussed. Finally, 
we demonstrate that we can obtain successful coupling constant unification in
models with leptoquarks, both with or without supersymmetry.  The supersymmetric
case requires the GUT group to be larger than $SU(5)$ such as flipped 
$SU(5)\times U(1)_X$.

\end{abstract}

\renewcommand{\thefootnote}{\arabic{footnote}} \end{titlepage} 

\section{Introduction}

The apparent symmetry between the quark and lepton generations is a mysterious
occurrence within the Standard Model (SM) and has inspired many theories which 
go beyond the SM to relate them at a more fundamental level.
As a result many of these models naturally contain leptoquarks, or particles
that couple to a lepton-quark pair.  Theories which fall in this category 
include, composite models with quark and lepton substructure\cite{comp}, the 
strong coupling version of the SM\cite{abfar}, horizontal symmetry 
theories\cite{sandip}, extended technicolor\cite{techni}, and grand unified 
theories (GUTs) based on the gauge groups $SU(5)$\cite{sufive}, $SO(10)$ with 
Pati-Salam $SU(4)$ color symmetry\cite{pati}, $SU(15)$\cite{framp}, and 
superstring-inspired $E_6$ models\cite{esix,phyrep}.
In all cases, the leptoquarks carry both baryon and lepton number and
are color triplets under SU(3)$_C$.  In models where baryon and lepton number
are separately conserved, which includes most of the above cases, 
leptoquarks can be light (of order the electroweak scale) and still avoid 
conflicts with rapid proton decay.  Their remaining properties, such as spin, 
weak iso-spin, electric charge, chirality of their fermionic couplings, and 
fermion number, depend on the structure of each specific model.  If 
leptoquarks were to exist we would clearly need to determine these properties 
in order to ascertain their origin.

An excess of events at large values of $Q^2$ have recently been 
reported\cite{expt} by both the H1 and ZEUS collaborations at HERA
in their neutral current Deep Inelastic Scattering (DIS) data.  ZEUS has
collected $20.1\inpb$ of integrated luminosity in $e^+p$ collisions and 
observes 5 events with $Q^2>15,000\gev^2$ with $x>0.45$ and $y>0.25$, where 
$x$ and $y$ are the usual DIS scattering variables, while expecting 2 events 
from the SM in this region.
H1 reports 7 events in the kinematic region $m=\sqrt{xs}>180\gev$ and $y>0.4$,
compared to a SM prediction of $1.83\pm 0.33$ with $14.19\pm 0.32\inpb$ of
integrated luminosity.  Clearly, the statistical sample is too small at
present to draw any conclusions and it is likely that this excess is
merely the result of a statistical fluctuation.  Another possibility is that
this discrepancy is the result of deviations from current parton distribution
parameterizations at large $x$.  This case, however, has been examined by
the H1 and ZEUS collaborations\cite{expt} and is found to be unlikely.  Also,
such large modifications in the parton densities would most likely result in
disagreement with the dijet data samples at the Tevatron\cite{cteq,kuhl}.
It is also possible that this HERA data might signal the first hint of physics
beyond the SM.  Such an excess in event rate at large $Q^2$ is a classic
signature for compositeness if the events show no specific kinematic 
structure. 
This scenario has recently been analyzed\cite{vb} in light of the HERA data,
with the result that an $eeqq$ contact interaction with a right-left helicity
structure (in order to avoid the constraints arising from Atomic Parity 
Violation data discussed below) and a scale of $\sim 3\tev$ is consistent 
with the data.  However, if
instead, the events cluster in $x$ while being isotropic in $y$, they would 
signal the production of a new particle.  While the reconstructed values 
of the mass ($m=\sqrt{xs}$) of such a hypothetical particle show some spread
between the two experiments, they are consistent within the evaluated errors,
yielding a central value in the approximate range $200-210\gev$.

If this excess of events turns out to be the resonant production of a new 
particle, we will need to examine the possible classes of new lepton-hadron 
interactions which could give rise to such a signature.  At present, there
are three leading scenarios of this type: (i) models with leptogluons, (ii)
explicit R-parity violating interactions in Supersymmetric theories, and (iii)
leptoquarks.  We now briefly discuss the first two cases, with the remainder
of the paper being devoted to the third.

Color octet partners of ordinary leptons are expected to exist in composite
models with colored preons\cite{preons}.  These particles, denoted as
leptogluons, are fermions, carry lepton number and couple directly to a 
lepton-gluon pair
with an undetermined strength.  This effective interaction may be written as
\begin{equation}
{\cal L}_{eff}={g_s\over\Lambda}\left[\lambda_L\bar L\sigma^{\mu\nu}
\ell_{8,L}^a+\lambda_R\bar\ell_{8,R}^a\sigma^{\mu\nu}e_R\right]G^a_{\mu\nu}
\ell + h.c.\,,
\end{equation}
where $G^a_{\mu\nu}$ is the gluon field strength tensor, $\Lambda$ is the
compositeness scale, $L$ represents the lepton doublet under $SU(2)_L$, 
and $\lambda_{L,R}$ parameterizes the unknown coupling.  We note
that this effective Lagrangian is non-renormalizable.  If the chiral symmetry
in these models is broken by QCD effects, the leptogluons are 
expected\cite{preons} to have masses or order $\alpha_s\Lambda$, and hence
could be as light as a few hundred GeV for compositeness scales in the TeV
range.  Leptogluons in this mass range would clearly reveal themselves in 
high-$Q^2$ DIS at HERA.  A $\sim 200\gev$ leptogluon
would also be copiously produced at hadron colliders.  In fact, as we will
see below, present Tevatron data most likely excludes leptogluons in this
mass region.

The second scenario for resonant new particle production listed above is 
that of supersymmetric theories with 
explicit R-parity violating interactions.  The most general gauge and
supersymmetry invariant superpotential (with minimal field content) 
contains the terms
\begin{equation}
\lambda_{ijk}L_iL_j\bar E_k+\lambda'_{ijk}L_iQ_j\bar D_k+\lambda''_{ijk}
\bar U_i\bar D_j\bar D_k\,,
\label{rparsp}
\end{equation}
where the first two terms violate lepton number ($L$), the third violates baryon
($B$) number, the $\lambda$'s are {\it a priori} unknown Yukawa coupling 
constants, and the $i,j,k$ are generational indices where $SU(2)_L$ invariance
demands that $i\neq j$.  In the Minimal Supersymmetric Standard Model (MSSM) 
a discrete symmetry matter parity (or R-parity) is applied to
prohibit all these dimension four $B$ and $L$ violating operators.  However, it
is sufficient to ensure only that the $B$ and $L$ violating terms do not
exist simultaneously in order to preserve nucleon stability.  In fact, a 
discrete anomaly free $Z_3$ symmetry, denoted as baryon parity, naturally
allows for the $L$ violating operators, while forbidding the $\Delta B\neq 0$ 
operators\cite{ross}.  The phenomenology of these models is strikingly different
than in the MSSM, as elementary couplings involving an odd number of
supersymmetric particles now exist.  This results in the possible single
production of super-partners and an unstable Lightest Supersymmetric
Particle.  At HERA, the second
$L$ violating term in Eq. \ref{rparsp} can mediate single squark production.
This possibility was first considered in Ref. \cite{jlhrpar}, and later
examined in detail in Ref. \cite{herbi}.  The relevant terms in the interaction
Lagrangian for this case are
\begin{equation}
{\cal L}=-\lambda'_{ijk}\left[ \tilde u^j_L\bar d^k_Re^i_L
+(\tilde d^k_R)^*(\bar e^i_L)^cu^j_L \right] +h.c.
\end{equation}
The requirement of $SU(2)_L$ invariance combined with the fact that
the positrons must be scattered off of valence quarks in order to account for 
the event excess at HERA, leaves us with only one possible scenario, the
production of charm or stop squarks via the first term above.  This case has
been recently examined\cite{herbitwo} in light of the data and will not be 
considered further here.  However, we note that such singly produced squarks
must also decay via their R-parity conserving interactions (\eg, $\tilde q\to
q+\chi^0$) at competitive rates, and hence events with these signatures must
also be observed. We will comment on some of the difficulties associated with 
GUT theories with R-parity violation below.

We now examine the third candidate scenario above, 
the existence of $200-210$ GeV leptoquarks, in detail.

\section{What is a Leptoquark?}

The interactions of leptoquarks can be described by an effective low-energy
Lagrangian.  The most general renormalizable $SU(3)_C\times SU(2)_L\times 
U(1)_Y$ 
invariant leptoquark-fermion interactions can be classified by their fermion
number, $F=3B+L$, and take the form\cite{brw}
\begin{equation}
{\cal L}  =  {\cal L}_{F=-2} + {\cal L}_{F=0} \,,
\end{equation}
with
\begin{eqnarray}
{\cal L}_{F=-2} & = & (g_{1L}\bar q^c_Li\tau_2\ell_L+g_{1R}\bar u^c_Re_R)S_1
+\tilde g_{1R}\bar d^c_Re_R\tilde S_1+g_{3L}\bar q^c_Li\tau_2\vec\tau
\ell_L\vec S_3 \nonumber\\
& & +(g_{2L}\bar d^c_R\gamma_\mu\ell_L+g_{2R}\bar q^c_L\gamma^\mu e_R)V_{2\mu}
+\tilde g_{2L}\bar u^c_R\gamma^\mu\ell_L\tilde V_{2\mu} + h.c. \,,\\
{\cal L}_{F=0} & = & (h_{2L}\bar u_R\ell_L+h_{2R}\bar q_Li\tau_2e_R)R_2
+\tilde h_{2L}\bar d_R\ell_L\tilde R_2+(h_{1L}\bar q_L\gamma^\mu\ell_L
+h_{1R}\bar d_R\gamma^\mu e_R)U_{1\mu} \nonumber\\
& & +\tilde h_{1R}\bar u_R\gamma^\mu e_R\tilde U_{1\mu}
+h_{3L}\bar q_L\vec\tau\gamma^\mu\ell_L\vec U_{3\mu} + h.c.
\nonumber
\label{efflag}
\end{eqnarray}
assuming the fermionic content of the SM. If the SM fermion content is 
augmented it is possible that new LQ interactions may arise. 
Here, $q_L$ and $\ell_L$ denote the $SU(2)_L$ quark and lepton doublets,
respectively, while $u_R\,, d_R$ and $e_R$ are the corresponding singlets.
The indices of the leptoquark fields indicate the dimension of their $SU(2)_L$
representation.  The subscripts of the coupling
constants label the lepton's chirality.  For simplicity, the color and
generational indices have been suppressed.  Since, in general, these couplings 
can be intergenerational, there is the possibility of large,
tree-level flavor changing neutral currents and flavor universality violations.
As discussed in the next section, this can be avoided by employing the 
constraint that a leptoquark couple only to a single generation.
We see that the leptoquark fermionic couplings are baryon and lepton number 
conserving, hence avoiding the conventional problems associated with 
rapid proton decay.  Note that the leptoquarks with
fermion number ($F$) of $-2$ ($S$ and $V$) couple to $\ell q$, while the $F=0$
leptoquarks ($R$ and $U$) have $\ell\bar q$ couplings.  Once this effective 
Lagrangian is specified, the gauge couplings of the leptoquarks are completely
determined.  Thus, only the strength of the leptoquark's fermionic Yukawa
couplings remain unknown.  For calculational purposes, these 
couplings are generally scaled to the electromagnetic coupling,
\begin{equation}
\lambda_{L,R} = e\lamt_{L,R}  \,,
\end{equation}
where $\lambda_{L,R}$ generically represent the $g_{iL,R}$ and $h_{iL,R}$.  
We note that in extended technicolor theories, the leptoquarks 
(denoted there as $P_3$) have couplings which are proportional to the 
masses of the quark-lepton pair.  In this case the $qeP_3$ coupling is clearly 
too small to be of interest to us here.

The quantum numbers and structure of the quark-lepton couplings are 
summarized in Table 1 for each leptoquark species.  For the couplings we list
the helicity, relative strength, and fermion pairs with which a particular
leptoquark may couple, using the convention $\overline{LQ}_{F=-2}\to\ell q$
and $\overline{LQ}_{F=0}\to\ell\bar q$.  The leptoquark's branching fraction 
into charged leptons, $B_\ell$, is also given.  The weak isospin structure is
denoted by the brackets.  Due to gauge invariance we would 
expect all the leptoquarks within a given
$SU(2)_L$ representation to be degenerate apart from loop corrections.  
For future reference we have
also listed the $SU(5)$ representations of lowest dimension within which the 
leptoquark can be embedded. Note that we have assumed the conventional 
assignments of the SM fermions to be in the $\overline{\mbox{\bf 5}}$ 
and {\bf 10} representations. 
Generally,  only a subset of these possible leptoquark states are
contained within a particular model.  For example, the scalar $S_{1L,R}$ is the
leptoquark present in superstring-inspired $E_6$ theories.  One exception is
the GUT based on $SU(15)$, which contains all 14 possible leptoquark states!

{\small
\begin{table}
\begin{center}
\begin{tabular}{l|ccrccc}
\hline\hline
 & Leptoquark \rule{0pt}{15pt} 
 & SU(5) Rep & & $Q$  & Coupling & $B_\ell$ \\ \hline
Scalars & & & & & & \\ \hline
$F=-2$ \rule{0pt}{15pt}& $S_{1L}$        
       & {\bf 5} & 
       & $1/3$ & $\lambda_L\, (e^+\bar u)$, 
$\lambda_L\, (\bar\nu\bar d)$ & $1/2$ \\
       & $S_{1R}$        
       & {\bf 5} & 
       & $1/3$ & $\lambda_R\, (e^+\bar u)$ & 1 \\
       & $\widetilde S_{1R}$ 
       & {\bf 45} & 
       & $4/3$ & $\lambda_R\, (e^+\bar d)$ & 1 \\[2ex]
       &         &        
       & & $4/3$ & $-\sqrt 2\lambda_L\, (e^+\bar d)$ & 1 \\[-4ex]
       & $S_{3L}$        
       & {\bf 45} & $\left\{\rule{0pt}{30pt}\right.$
       & $1/3$& $-\lambda_L\, (e^+\bar u)$,
$-\lambda_L\, (\bar\nu\bar d)$ & $1/2$ \\[-4ex]
       &         &        
       &         & $-2/3$& $\sqrt 2\lambda_L\, (\bar\nu\bar u)$ & 0 \\[2ex]
\raisebox{-2ex}[0pt]{$F=0$} &  
\raisebox{-2ex}[0pt]{$R_{2L}$} &        
\raisebox{-2ex}[0pt]{{\bf 45}} &
\raisebox{-2ex}[0pt]{$\left\{\rule{0pt}{20pt}\right.$} 
& $5/3$ & $\lambda_L\, (e^+u)$ & 1 \\[-4ex]
       &  &  &        & $2/3$ & $\lambda_L\, (\bar\nu u)$ & 0 \\[2ex]
       & 
\raisebox{-2ex}[0pt]{$R_{2R}$} &
\raisebox{-2ex}[0pt]{{\bf 45}} & 
\raisebox{-2ex}[0pt]{$\left\{\rule{0pt}{20pt}\right.$}
& $5/3$ & $\lambda_R\, (e^+u)$ & 1 \\[-4ex]
  & &    &  & $2/3$ & $-\lambda_R\, (e^+d)$ &  1\\[2ex]
       & 
\raisebox{-2ex}[0pt]{$\widetilde R_{2L}$} & 
\raisebox{-2ex}[0pt]{{\bf 10/15}} & 
\raisebox{-2ex}[0pt]{$\left\{\rule{0pt}{20pt}\right.$} &
$2/3$ &$\lambda_L\, (e^+d)$ & 1 \\[-4ex] 
       &       &    &         & $-1/3$& $\lambda_L\, (\bar\nu d)$ 
&  0 \\[1.2ex] \hline
Vectors & & & & & \\ \hline
\rule{0pt}{15pt}
\raisebox{-3ex}[0pt]{$F=-2$} & 
\raisebox{-2ex}[0pt]{$V_{2L}$} & 
\raisebox{-2ex}[0pt]{{\bf 24}} & 
\raisebox{-2ex}[0pt]{$\left\{\rule{0pt}{20pt}\right.$} &
 $4/3$ & $\lambda_L\, (e^+\bar d)$ & 1 \\[-4ex]
       & & && $1/3$ & $\lambda_L\, (\bar\nu\bar d)$ & 0 \\[2ex]
       & 
\raisebox{-2ex}[0pt]{$V_{2R}$} & 
\raisebox{-2ex}[0pt]{{\bf 24}} &
\raisebox{-2ex}[0pt]{$\left\{\rule{0pt}{20pt}\right.$} & 
$4/3$ & $\lambda_R\, (e^+\bar d)$ & 1 \\[-4ex]
       &   &              &         &
        $1/3$ & $\lambda_R\, (e^+\bar u)$ & 1 \\[2ex]
       & 
\raisebox{-2ex}[0pt]{$\widetilde V_{2L}$} & 
\raisebox{-2ex}[0pt]{{\bf 10/15}} &
\raisebox{-2ex}[0pt]{$\left\{\rule{0pt}{20pt}\right.$} &
 $1/3$ & $\lambda_L\, (e^+\bar u)$ & 1 \\[-4ex]
       &       &          &         &
        $-2/3$ & $\lambda_L\, (\bar\nu\bar u)$ & 0 \\[2ex]
$F=0$  & $U_{1L}$  & {\bf 10} && $2/3$ & $\lambda_L\, (e^+d)$,
$\lambda_L\, (\bar\nu u)$ & $1/2$ \\
       & 
$U_{1R}$  & {\bf 10} &&  $2/3$ & $\lambda_R\, (e^+\bar d)$ & 1 \\
       & 
$\widetilde U_{1R}$ & 
{\bf 75} &&
 $5/3$ & $\lambda_R\, (e^+u)$ & 1 \\[2ex]
       &    &             &         &
        $5/3$ & $\sqrt 2\lambda_L\, (e^+u)$ & 1 \\[-4ex]
       & 
$U_{3L}$  & 
{\bf 40} &
$\left\{\rule{0pt}{30pt}\right.$ &
 $2/3$ & $-\lambda_L\, (e^+d)$, $\lambda_L\, (\bar\nu u)$ & $1/2$\\[-4ex]
       &      &           &         &
        $-1/3$ & $\sqrt 2\lambda_L\, (\bar\nu d)$ & 0 \\[2ex]
\hline\hline
\end{tabular}
\caption{Quantum numbers and fermionic coupling of the leptoquark states.  No
distinction is made between the representation and its conjugate.}
\end{center}
\label{lqprop}
\end{table}}

Using Eq. 5, the total leptoquark tree-level decay 
widths are easily calculated to be
\begin{equation}
\Gamma = \begin{array}{cc}
{\mbox{$\alpha m_{LQ}$}\over \mbox{$4$}} \sum_i \lamt_i^2 = 0.386\,
{\mbox{$m$}\over \mbox{$200 \gev$}}
\sum_i \lamt_i^2 \gev\,, & {\mathrm {Scalars}} \\
{\mbox{$\alpha m_{LQ}$}\over \mbox{$6$}} \sum_i \lamt_i^2 = 0.258\,
{\mbox{$m$}\over \mbox{$200 \gev$}}
\sum_i \lamt_i^2\gev\,, & {\mathrm {Vectors}}
\end{array}
\label{width}
\end{equation} 
where we have scaled the widths to a $200\gev$ leptoquark in our numerical
evaluation, and the sum extends over all possible decay modes.
These states are clearly very narrow and hence are long-lived, especially for
the values of $\lamt$ that are consistent with the low-energy constraints 
discussed in the next section. As pointed out by Kunszt and Stirling, QCD 
corrections to these widths are very small{\cite {ks}}. 

\section{Low Energy Constraints}

As mentioned above, low-energy data places strong restrictions on the
leptoquark Yukawa couplings.  In this section we summarize the most relevant
of these constraints.

As is well known, for a leptoquark to be sufficiently light for it to be of 
phenomenological interest at existing or planned colliders,  it must have
essentially chiral couplings to fermions.  Here, we give
two examples which demonstrate this conclusion.  ($i$) Consider 
a first generation leptoquark coupling to $ue^-_R$ with strength $\lambda_R$,
and to $d\nu_L$ with strength $\lambda_L$. A Fiertz transformation then yields 
the interaction 
\begin{equation}
{\cal L} = {\lambda_L \lambda_R \over 2m^2_{LQ}}\bar u_Rd_L\bar e_R\nu_L \,,
\end{equation}
which gives a large contribution to the decay $\pi^+\to e^+\bar\nu_e$.
Comparison with current data results\cite{dbc,leurer} in the bound
$m_{LQ}>200|\lambda_L \lambda_R|^{1/2}$ TeV, implying that at least one if not 
both of these Yukawa couplings must be small, if leptoquarks are to be light.  
If one of these couplings is 
sufficiently large in order to induce leptoquark-fermion interactions at an
interesting level, then the coupling with the other handedness must essentially
vanish, \ie, the couplings are chiral. ($ii$) One can also
examine the leptoquark's contribution to the $g-2$ of the muon arising from 
a one loop penguin diagram involving a light quark and a leptoquark. 
In this case if couplings of both 
helicities are present, current data places the constraint\cite{dbc}
$M_{LQ}>1000\lambda_L \lambda_R$ TeV, and again we see that these couplings
must be essentially chiral.

Even if the leptoquark-fermion couplings are chiral, strong constraints on their
magnitude still arise from their potential contributions to 
a wide class of flavor changing neutral 
currents and related phenomena.  An exhaustive study of this class of
transitions has been performed by Davidson, Bailey and Campbell{\cite {dbc}},
and hence we will not repeat this type of investigation here.  The results
of this study show that most of the stronger bounds can be 
satisfied if the requirement that a given leptoquark couple to only one 
generation is imposed. This results in the nomenclature of first, second 
and third generation leptoquarks found in the literature.  Note that this 
restriction is more severe than the simple requirement of family number 
conservation at a single vertex.  This is easily illustrated by examining the 
process $K\to \mu e$,
which would receive a large tree-level contribution if leptoquarks
were allowed to simultaneously couple to both the first and second families.

After the constraints of chiral and single generation couplings are imposed,
there are two important
remaining low-energy constraints arising from Atomic Parity Violation (APV) and 
the universality testing decay $\pi \to e\nu$. Comparable but somewhat weaker 
bounds also follow from quark-lepton universality. 
These additional restrictions have been examined by Leurer{\cite {leurer}} and 
also by Davidson \etal\cite{dbc} for both cases of
scalar and vector leptoquarks.  We summarize these results in Table 2
(assuming that the LQ is not responsible for the small difference between the 
SM expectations for the APV `weak charge' and what is obtained experimentally). 
These values have now been updated to 
include the recent results of Wood \etal on APV in Cesium{\cite {wood}} 
which are in good agreement with SM predictions{\cite {ros}} yielding 
$\Delta Q_W=1.09\pm 0.93$. 
Note that these bounds are far from trivial. For example, 
we see from this Table that a spin-0, $\widetilde R_{2L}$ leptoquark 
with a mass of 200 GeV must have $\tilde \lambda <0.22$! As we 
will see below this is not far from the value suggested by the excess of
events at HERA. 

\begin{table}
\begin{center}
\begin{tabular}{lc|lc}
\hline\hline
Leptoquark  & Limit  & Leptoquark & Limit \\
\hline
$S_{1L}$          & 1040 & $U_{1L}$        & 1300 \\
$S_{1R}$          & 833  & $U_{1R}$        & 645  \\    
$\widetilde S_{1R}$   & 901  & $\widetilde U_{1R}$ & 597  \\
$S_{3L}$          & 695  & $U_{3L}$        &1993 \\
$R_{2L}$          & 572  & $V_{2L}$        & 1605  \\
$R_{2R}$          & 833  & $V_{2R}$        & 645 \\
$\widetilde R_{2L}$   & 901  & $\widetilde V_{2L}$ &597 \\
\hline\hline
\end{tabular}
\caption{Combined limits on the ratio $m/\tilde\lambda$ in GeV, where $m$
is the leptoquark mass, for the
leptoquarks multiplets from data on Atomic Parity Violation and the decay 
$\pi \to e\nu$. }
\end{center}
\label{lowe}
\end{table}

Other types of experiments give slightly weaker bounds on the Yukawa
couplings for fixed leptoquark mass. For example, it is well-known that 
precision measurements in deep inelastic neutrino scattering are
sensitive to new particle exchanges. Using the latest CCFR results{\cite{ccfr}}
we obtain the constraint $\tilde \lambda_{iL,R}\lsim 0.4-0.6$ for 200 GeV
leptoquarks.  NuTeV\cite{ccfr} may be able to improve this reach by
a factor of two.  Older results, such as that from the SLAC polarized 
electron-Deuteron scattering experiment{\cite {prescott}}, 
are only sensitive to $\tilde \lambda_{iL,R}$ of order unity or greater.

\section{Leptoquarks at HERA}

Clearly, ep collisions are especially well suited for leptoquark production 
and offer striking signals\cite{brw,epprod,epasym}.  
Direct production contributes to DIS in either
the neutral or charged current channel, through an s-channel
resonance $e^\pm~^(\bar q^)\to LQ$ with the subsequent decay to either
$e^\pm~^(\bar q^)$ or $\nu_e~^(\bar q^)$ with a fixed branching fraction, 
depending on the leptoquark species.  The final state with $\nu_e$ manifests 
itself as missing energy.  Clearly, this s-channel exchange would yield
distinctive, and due to the size of the width, narrow peaks in the 
$x$-distributions at $x=m^2/s$.  These peaks, however, are smeared by the 
detector resolution as well as QCD and QED radiative 
effects\cite{harnew,lund}.  
Additional smaller contributions are generated from 
u-channel leptoquark exchange.  The fermion number of the leptoquark dictates
whether it will contribute via s- or u-channel exchange in $e^-$ versus
$e^+$ scattering off of valence ($q$) or sea ($\bar q$) partons.  For example,
the $F=-2$ leptoquarks ($S$ and $V$) mediate DIS through the s-channel 
(u-channel) in $e^+\bar q$ ($e^+q$) collisions, while the $F=0$ states are
exchanged in the u-channel (s-channel) in $e^+\bar q$ ($e^+q$) collisions.
In principle, this can be used to separate the production of $F=-2$ from $F=0$ 
leptoquarks from cross section measurements alone.

We first examine the expected event yield in the $e^\pm j$ channel at HERA 
for the production of the various leptoquark species.  We concentrate on the
case of scalar leptoquarks, as $\sim 200\gev$ vector leptoquarks are most
likely excluded by Tevatron data as shown in the next section.  
The cross section is dominated by the s-channel resonance, and the narrow width
of these states as seen as Eq. \ref{width} justifies the use of the 
narrow width approximation.  The differential cross section for leptoquark
production can then be written as
\begin{equation}
{d\sigma(ep\to LQ\to ej)\over dy} = 
{\pi^2\alpha\over s}\lamt^2~^(\bar q^)(m^2/s,Q^2)B_\ell
\left\{ \begin{array}{cc}
1\,, & {\mathrm{Scalar}} \\
6(1-y)^2\,, & {\mathrm{Vector}} 
\end{array} \right\} \,.
\end{equation}
For the case of scalar leptoquarks, we see that the production is isotropic in 
y, whereas the electroweak DIS background has a $1/y^2$ behavior.  In
obtaining the total cross section for scalar leptoquarks, 
we use the y-averaged parton densities
\begin{equation}
\langle q(x,sxy)\rangle = {\int q(x,sxy)dy\over\int dy} \,,
\end{equation}
with the range $0.25\leq y\leq 1.0$, and employ the MRSA$'$ 
distributions\cite{mrsa}.  The resulting expected excess of events from
scalar leptoquark production are displayed in Table \ref{lqprod} 
for $e^+p$ and $e^-p$ collisions, scaled
to $20\inpb$ and $1\inpb$ of integrated luminosity, respectively.  Here we
assume $m=200\gev$ and for purposes of demonstration we take
$\lamt=0.1$.  The numbers
in brackets represent the corresponding event rate in the $\slash p_Tj$ channel.
Note that only 2 leptoquark species can contribute to charged current DIS 
assuming only the SM fermion content as shown in Eq. 5.
We stress that the relative magnitudes of these event rates are fixed by
the contributing parton densities and the luminosity.  As expected, the $S_i$ 
leptoquarks have significantly larger cross sections in electron (rather than 
positron) collisions, since the valence quark distributions contribute in this 
case.  Thus, in order to account for the HERA data, the $F=-2$ scalar 
leptoquarks would also yield an excess of events in the $\sim 1\inpb$ of
$e^-p$ data.  For example, if the leptoquark contained in $E_6$ theories, 
$S_{1L}$, were to account for the observed $e^+j$ excess, then it would also
yield roughly 600 excess events in $1\inpb$ of $e^-p$ data!  Thus, unless
there are significant event excesses hiding in the H1 and ZEUS $e^-p$ data
(which is not yet completely analyzed), we may exclude $F=-2$ S-type 
leptoquarks as the source of the HERA events.
In contrast, $F=0$ scalar leptoquark production is suppressed in electron
collisions.  We see that in the case of $\widetilde R_{2L}$, the predicted 
number of events in the $e^+j$ channel with the assumed coupling strength of 
$\lamt\simeq0.1$ is consistent with the data, while for $R_{2L,R}$ the coupling 
would have to be somewhat smaller with $\lamt\sim 0.03-0.04$ neglecting the
potentially large QCD corrections. 

\begin{table}
\centering
\begin{tabular}{ccc} \hline\hline
Leptoquark& $N_{e^+} (20\inpb)$ & $N_{e^-} (1\inpb)$ \\
\hline

$S_{1L}$        & 0.054 [0.054]& 0.591 [0.591] \\
$S_{1R}$        & 0.108 & 1.18 \\
$\widetilde S_{1R}$ & 0.229 & 0.288 \\
$S_{3L}$        & 0.512 [0.054] & 1.17 [0.591] \\
$R_{2L}$        & 23.7 & 0.005 \\
$R_{2R}$        & 29.3 & 0.017 \\
$\widetilde R_{2L}$ & 5.58 & 0.012\\
\hline\hline
\end{tabular}
\caption{Number of events for scalar leptoquark production in $e^+p$ and $e^-p$
collisions, scaled to $20\inpb$ and $1\inpb$ of integrated luminosity,
respectively, assuming $m=200$ GeV, $\lambda=0.1 e$, and taking $0.25\leq y
\leq 1.0$.  The numbers in brackets indicate the corresponding expected
event yield in the charged current channel.}
\label{lqprod}
\end{table}

If the leptoquark signature is
verified by future data taking at HERA, it will be mandatory to determine
its couplings.  Clearly, the best method of accomplishing this at HERA is to
use both $e^\pm p$ collisions and to take advantage of possible beam
polarization\cite{brw,epasym}.  
(It is expected that polarization levels of $P\approx 50\%$
may be achievable at HERA in the future.)  Table \ref{herarates} displays the
total number of expected events, scaled to $100\inpb$, assuming $100\%$ beam 
polarization for $e^\pm_{L,R}p$
collisions for a 200 GeV scalar leptoquark of each type, subject to the cuts
$0.4\leq y\leq 1.0$ and $M_{ej}=200\pm 20\gev$ to remove the SM background. 
In these calculations the full deep inelastic scattering amplitudes, including 
the exchanges in all channels, have been used. The numerical results justify 
our earlier use of the narrow width approximation. 
The results have also been smeared by a $5\%$ mass resolution.  It is clear
from the Table that knowledge of the {\it ratio} of cross sections which are 
essentially free of QCD corrections in the four
channels will allow the leptoquark quantum numbers to be determined if
sufficient statistics are available. For fixed values of $\lambda$, it has 
been shown{\cite {ks,pssz}} that the QCD corrections to the production of 
leptoquarks off of the valence partons can be as large as $+25\%$, but are 
somewhat smaller for the case of production off of sea quarks. 

\begin{table}
\begin{center}
\begin{tabular}{lcccc}
\hline\hline
Leptoquark&$N_L^-$ & $N_R^-$ & $N_L^+$ & $N_R^+$ \\
\hline
SM background   & 51.7 & 28.7 & 9.98 & 20.0 \\
$S_{1L}$        & 121. & 28.7 & 9.98 & 20.4 \\
$S_{1R}$        & 51.7 & 167. & 10.8 & 20.0 \\
$\widetilde S_{1R}$ & 51.7 & 63.0 & 11.5 & 20.0 \\
$S_{3L}$        & 190. & 28.7 & 9.98 & 23.5 \\
$R_{2L}$        & 52.4 & 28.7 & 9.98 & 158. \\
$R_{2R}$        & 51.7 & 29.4 & 148. & 20.0 \\
$\widetilde R_{2L}$ & 53.2 & 28.7 & 9.98 & 54.4 \\
\hline\hline
\end{tabular}
\caption{Number of events per $100^{-1}pb$ for each electron charge and 
state of polarization for a 200 GeV scalar leptoquark at HERA assuming 
$0.4<y<1$, $\tilde \lambda=0.1$, and $M_{ej}=200\pm 20$ GeV. These results 
have been smeared with a detector resolution of $5\%$ in $M_{ej}$.}
\end{center}
\label{herarates}
\end{table}

\section{Signatures at the Tevatron}

Leptoquarks may reveal themselves in several reactions at hadron colliders.
They may be observed directly via pair or single production mechanisms or
they may indirectly influence the lepton pair invariant mass spectrum in
Drell-Yan processes.  We examine each of these in this section.

\begin{figure}[htbp]
\centerline{
\psfig{figure=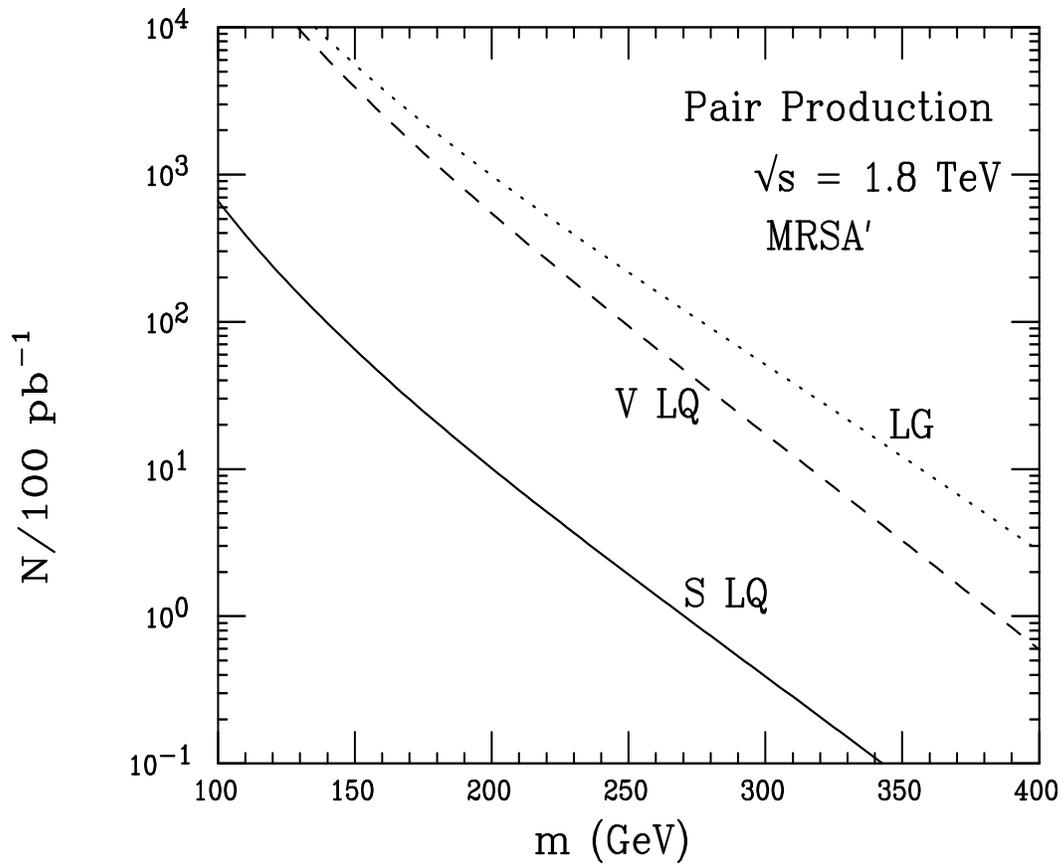,height=14.cm,width=16cm,angle=-90}}
\vspace*{-0.75cm}
\caption{Pair production cross sections at the Tevatron for scalar
and vector leptoquarks, as well as leptogluons, corresponding to the solid,
dashed, and dotted curves, respectively.  Here, the $gg$ and $q\bar q$ 
contributions are summed.}
\label{lqpairs}
\end{figure}

Since leptoquarks are color triplet particles, their pair 
production\cite{jlhsp,blum}
proceeds through gluon fusion or quark annihilation and is essentially
{\it independent} of the Yukawa coupling, $\lambda$.  There is a
potential contribution of order $\lambda^2$ via the reaction 
$q\bar q\to LQ\overline {LQ}$ with t-channel lepton exchange, however this 
contribution is negligible for the size of Yukawa couplings of relevance here,
$\lambda\sim{\cal O}(10^{-1}e)$. The pair production of scalar leptoquarks thus 
mimics that of squarks.  The number of events expected at the Tevatron, scaled
to $100$ pb$^{-1}$ of integrated luminosity, for the pair production of
one generation of a single type of
scalar leptoquark is displayed as the solid curve in Fig. \ref{lqpairs}.
Here, we have employed the MRSA$'$ parton distributions\cite{mrsa}, and omitted 
the $K$-factor which has been  calculated\cite{kfac} for
leptoquark production to be $K_{gg}=1+2\alpha_s\pi/3$ and $K_{qq}=1-\alpha_s
\pi/6$ for the gluon fusion and quark annihilation subprocesses, respectively. 
For a 200 GeV scalar leptoquark, this yields an enhancement in the cross section
by a factor of 1.16 giving $\sigma=0.117$ pb assuming $\mu^2=\hat s$. If
instead $\mu^2=m^2$ is chosen, a larger cross section will result since we 
always have $\hat s > 4m^2$. We note that the cross section
falls rapidly, dropping by a factor of 1.32 between $m=200$ and $210\gev$  with 
$\sigma(m=210\gev)=0.089$ pb (including the K-factor). We note that a complete 
NLO calculation of scalar LQ pair production has now been completed by 
Kr\"amer \etal which essentially reproduces the results obtained using the 
K-factor approach at a scale of $\mu^2=m^2${\cite {kra}}. 

The signatures for leptoquark pair production are 2 jets
accompanied by either $\ell^+\ell^-$, $\ell^\pm\slash p_T~$ or $\slash p_T~$,
with a pair of jet$+\ell^{\pm,0}$ invariant masses being equal to the mass
of the leptoquark.  D0 has searched for the dijet with two or single charged
lepton topologies in the electron and muon channels,
and CDF has searched in the dilepton+dijet case for all three lepton
generations.  For each generational coupling the most stringent 
bounds\cite{tevlq} are $m_{(e)}>175 (147)$ GeV from D0, $m_{(\mu)}>180 (140)$
GeV and $m_{(\tau)}>99$ GeV from CDF, with $B_\ell=1 (0.5)$.  D0 sets a $95\%$ 
C.L. limit on the pair production cross section of scalar leptoquarks decaying
into $\epem jj$ of $\sigma\lsim 0.25$ pb$^{-1}$.  
This D0 bound is based on the observation of 3 events with a Monte Carlo 
background estimate of $2.85\pm 1.08$ events arising from Drell-Yan, 
$t\bar t \to \ell \ell$, $Z\to \tau \tau \to \ell \ell$, $WW \to \ell \ell$ 
and fakes from QCD. Clearly the leptoquark $SU(2)_L$ 
representations which contribute to the cross 
section significantly more than that of a single leptoquark, \eg, $R_{2R}$, 
will have a more difficult time satisfying these bounds. 
We stress again that these constraints
are independent of the value of $\lambda$.

CDF has yet to present results from a search for first generation leptoquarks
from runs 1A and 1B, which contain a combined data sample of approximately 
$110 pb^{-1}$.  Based on our cross section above, we would expect CDF to 
observe at most 1 signal event (taking $B_\ell=1$) depending exactly on 
the leptoquark mass (\eg, 200 versus 210 GeV) and the 
details of their selection criterion, plus an unknown amount of background. 
Until all of these numbers become available we cannot attempt to combine the 
CDF and D0 results to obtain a stronger mass bound without speculating upon the 
details of the CDF data and a thorough understanding of the common 
systematics of the two experiments.  However, we would not expect a combined
D0/CDF bound to significantly exceed 200 GeV.

In order to compute the pair production cross section for vector leptoquarks
($V$) we need to determine both the trilinear $gVV$ and quartic $ggVV$
couplings.  In any realistic model that contains fundamental vector leptoquarks,
they will be the gauge bosons of some extended gauge group.  Hence gauge 
invariance will completely specify the $gVV$ and $ggVV$ couplings in
such a manner as to guarantee that the subprocess cross section obeys
tree-level unitarity, as is the hallmark of all gauge theories.  However,
vector leptoquarks could be a low-energy manifestation of a more fundamental
theory at a higher scale, and they could be composite.  In this case various
anomalous $gVV$ and $ggVV$ couplings could be present, one of which can be 
described by a chromo-magnetic moment, $\kappa$.  This term 
represents the only dimension 4 anomalous coupling which conserves CP.  In 
gauge theories $\kappa$ takes on the value of unity.  The cross sections for
vector leptoquarks with $\kappa=1$ have been computed in 
Refs. \cite{blum,hrphp,oscar}.  In this case the total event 
rate, scaled again to $100\inpb$ of integrated luminosity, for $VV$ production 
at the Tevatron is given by the dashed curve in Fig. \ref{lqpairs}.
Vector leptoquarks have the same signatures as discussed above for
the scalar case, but with slight detailed variations in the production
angular distribution due
to the fact they are spin-1 particles.  A reasonable estimate of the search
reach can be obtained by employing the D0 bound on the cross section for
$\epem jj$ events from scalar leptoquark production.  We estimate that this 
procedure yields
the constraint on first generation vector leptoquarks with $\kappa=1$ of 
$m\gsim 290$ GeV, placing this case out of the kinematic reach of 
HERA{\cite {hag}}.
Clearly, the experiments themselves need to perform a detailed analysis in
order to confirm this estimate.

\begin{figure}[htbp]
\centerline{
\psfig{figure=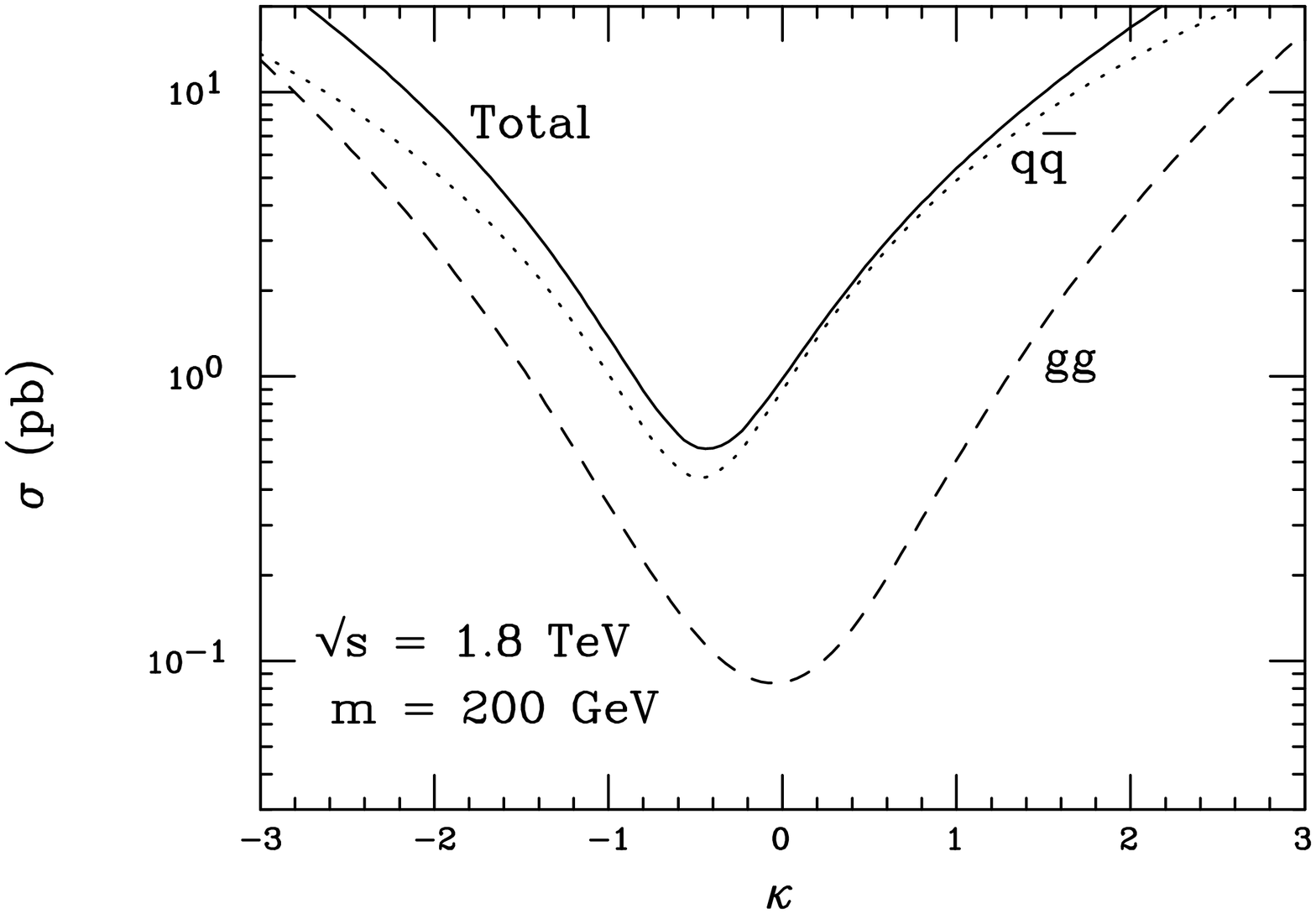,height=14.cm,width=16cm,angle=-90}}
\vspace*{-0.75cm}
\caption{$\kappa$ dependence of the $q\bar q$, $gg$, and total
vector leptoquark pair production cross sections at the Tevatron, represented 
by the dotted, dashed, and solid curves, respectively.  The leptoquark mass is
taken to be 200 GeV.}
\label{vvkap}
\end{figure}

The results in the more general case\cite{blum,hrphp} of $\kappa\neq 0$ are 
displayed in Fig. \ref{vvkap}.  
Here, the separate $q\bar q$, $gg$, and the total cross 
sections for vector leptoquark pair production with $m=200$ GeV are shown 
as a function of $\kappa$.  We see that the cross section varies significantly
with $\kappa$ yielding larger or smaller values than the results given above
for $\kappa=1$.  In the worst case, the total cross section reaches its lowest
value of $\sigma_{low}\approx 0.6$ pb around $\kappa\approx -0.45$.  
This value of $\kappa$ is, of course, much larger than one would expect in
any realistic model but is considered here for generality.  However, this value
of $\sigma_{low}$ is significantly 
larger than the D0 bound\cite{tevlq} of $0.25$ pb and 
would again exclude 200 GeV vector leptoquarks for this 
$\kappa$ with $B_\ell=1$.  For all 
other values of $\kappa$, the cross section is comfortably large enough to
be prohibited by D0.  We remind the reader
that we have neglected the K-factors in this analysis; their inclusion 
will only strengthen our conclusions.

Here, we also consider the case of pair production of leptogluons at the
Tevatron.  This process was considered in Refs. \cite{tgrlg,ulilg} and is also 
mediated by gg fusion and $q\bar q$ annihilation, similar to 
the production of any heavy colored fermion.  However,
since leptogluons are color octets, there is an enhanced color structure in
this case compared to, \eg,  top-quark pair production.  The event rate,
scaled to $100\inpb$, is given by the dotted curve in Fig. \ref{lqpairs}.
We see that this cross section is larger than that for both scalar and vector
leptoquarks, and is roughly six times the top pair cross section.  
The D0 cross section bounds on $\epem jj$ events would clearly exclude 200
GeV leptogluons and could naively place the constraint $m_{LG}\gsim 325$ GeV.

\vspace*{-0.5cm}
\nn

\vspace*{-0.5cm}
\nn

\begin{figure}[htbp]
\centerline{
\psfig{figure=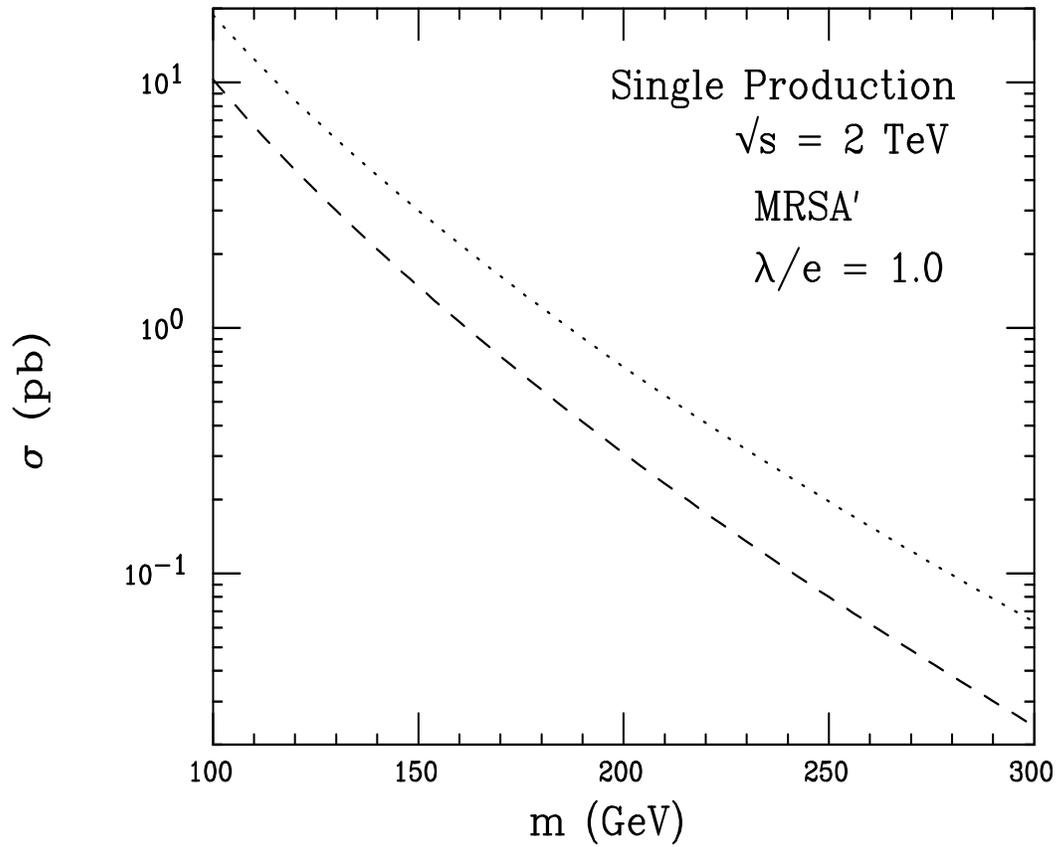,height=14.cm,width=16cm,angle=-90}}
\vspace*{-0.75cm}
\caption{Single production cross sections at the Tevatron for scalar
leptoquarks as a function of mass from $gu(\bar u)$ and $gd(\bar d)$ fusion, 
corresponding to the dotted and dashed curves, respectively.  Here, 
$\lambda/e=1$, and the K-factor has been omitted.}
\label{lqsing}
\end{figure}

Leptoquarks can also be produced singly  at hadron colliders.
The parton level subprocess responsible for single production is $qg\to 
LQ+\ell$, where $\ell$ is a charged lepton or neutrino depending on the type of 
leptoquark.  The diagrams for this process contain the QCD strong coupling
at one vertex and the leptoquark Yukawa coupling at the other.  The subprocess
differential cross section for scalar leptoquarks is given by\cite{jlhsp,singlq}
\begin{equation}
{d\hat\sigma\over d\that} = {\pi\alpha_s\alpha\lamt^2\over 3\shat^2}
\left[{\shat+\that-m^2\over\shat}+{\that(\that+m^2)\over (\that-m^2)^2}
+{\that(2m^2-\shat)\over\shat(\that-m^2)}\right]\,,
\end{equation}
where $\that={1\over 2}(m^2-\shat)(1-\cos\theta)$ with $\theta$ being the
quark-lepton scattering angle.  For completeness we give the corresponding
single production cross section for vector leptoquarks in the Appendix.
Compared to pair production, this mechanism has the advantage of a larger
amount of available phase space, but has the disadvantage in that it is directly
proportional to the small Yukawa coupling.  The total cross section at the 
Tevatron Main Injector and/or TeV33(now taking $\sqrt s=2$ TeV) in the case of 
scalar leptoquarks with 
$\lambda/e=1$ for both $gu+g\bar u$ and $gd+g\bar d$ fusion is presented in 
Fig. \ref{lqsing}.  Note that for a $\bar pp$ collider, the $gq$ and $g\bar q$
cross sections are equal.  Here, we have again omitted the K-factor 
(given by $K=1+3\alpha_s\pi/4$ for single production\cite{kfac}), which enhances
the cross section by a factor of $\simeq 1.25$.
For a 200 GeV scalar leptoquark with coupling strength of 
$\lambda/e=0.1$ and including the K-factor, our results show that we obtain 
approximately $\sim 39\,, 87$ events from 
$gd+g\bar d\,, gu+g\bar u$ fusion,  respectively, would be obtained with $10$ 
fb$^{-1}$ of integrated luminosity at the Main Injector/TeV33.  This event 
rate should be sufficient to provide a very rough determination of the 
value of Yukawa coupling $\lambda$.
The signatures for this production mode are jet $+\ell^+\ell^-\,, 
+\ell^\pm\slash p_T\,, +\slash p_T$ and at least in the first case, should be
easily detectable.

\vspace*{-0.5cm}
\nn

Indirect signals for leptoquarks may be observed in Drell-Yan 
production\cite{dryan}.
In addition to the usual s-channel $\gamma$ and $Z$ exchange in the SM,
leptoquarks may also contribute to $q\bar q\to\epem$ in t- or u-channel
exchange, with the specific channel depending on the leptoquark type, via the 
Yukawa coupling.  This new exchange will modify not only the invariant mass 
distribution, but also the angular distribution of the lepton pair.  The
parton level differential cross section for this process can be written as
\begin{equation}
{d\sigma\over dMdydz} = {K\pi\alpha^2\over 6M^3} \sum_q[F_q^+G_q^+ + F_q^-G_q^-]
\,,
\label{dyan}
\end{equation}
where $M$ represents the invariant mass of the lepton pair,
$z=\cos\theta^*$ with $\theta^*$ being the $\ell^+\ell^-$ center-of-mass
scattering angle, $K$ is the usual QCD correction factor with $\alpha_s$
evaluated at the scale $M$, and the sum extends over the appropriate partons.
In the SM all quark flavors, in principle, contribute to this process,
whereas, in the case of leptoquark exchange, only one or two quark flavors 
contribute.
The parton density factors are given by
\begin{equation}
G_q^\pm=x_ax_b[q(x_a)\bar q(x_b)\pm q(x_b)\bar q(x_a)] \,,
\end{equation}
with $x_{a,b}=(M/{\sqrt s})e^{\pm y}$ as usual, and $F_q^\pm$ represent 
the even and odd kinematic functions, which are given in the Appendix for
both the SM and leptoquark contributions.

\begin{figure}[htbp]
\centerline{
\psfig{figure=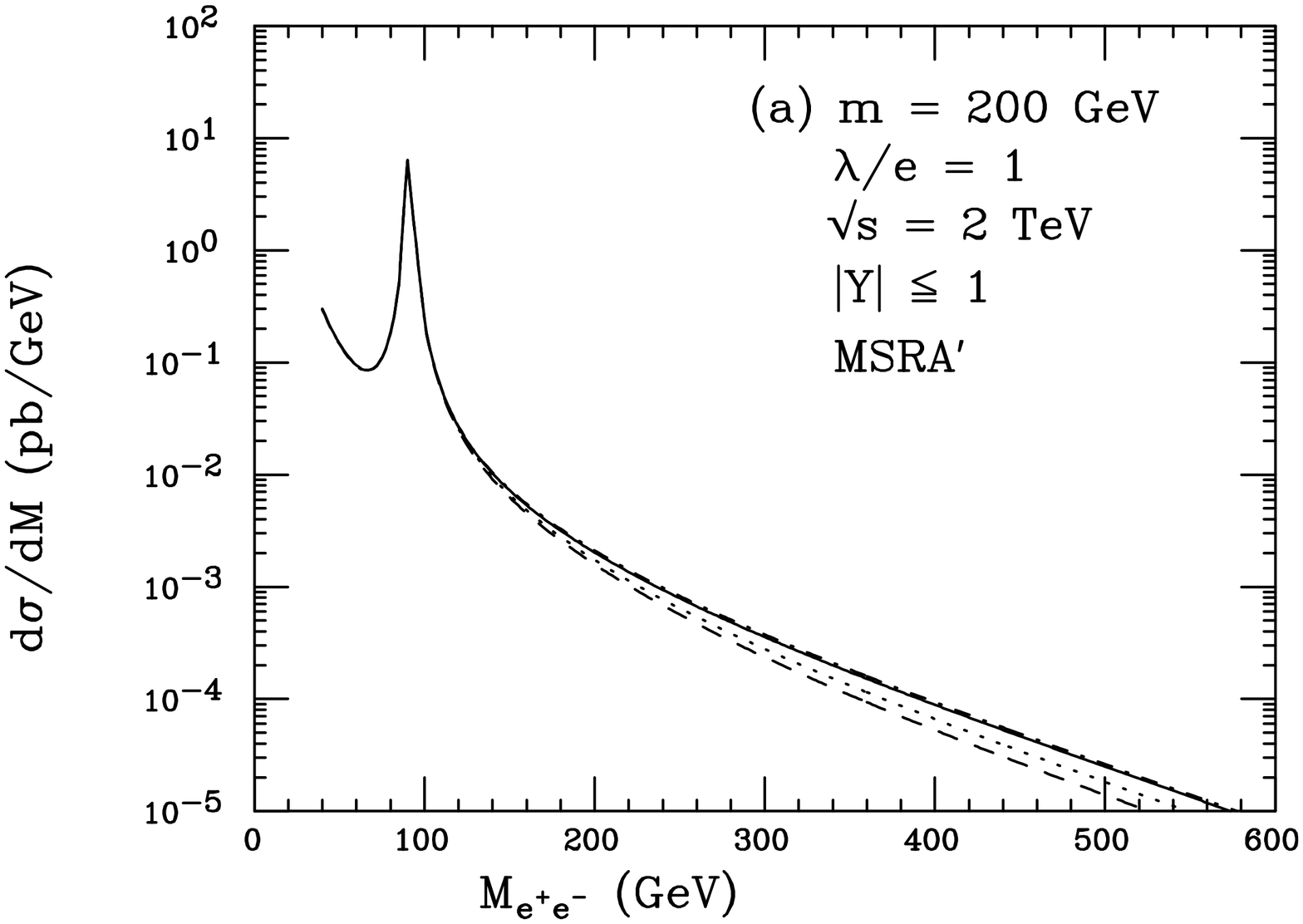,height=10.cm,width=12cm,angle=-90}}
\vspace*{-5mm}
\centerline{
\psfig{figure=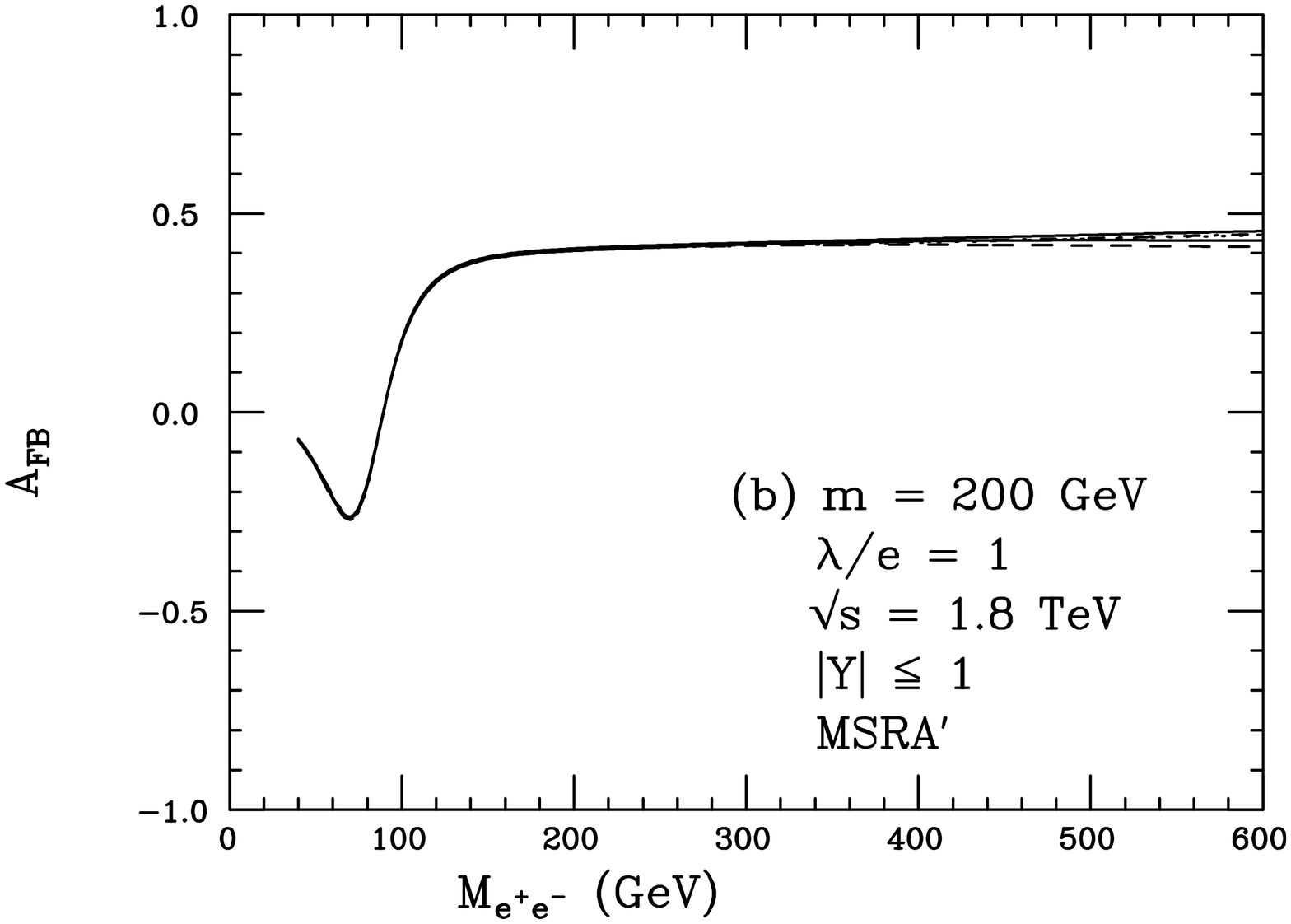,height=10.cm,width=12cm,angle=-90}}
\vspace*{-0.75cm}
\caption{(a) The lepton pair invariant mass distribution and (b) 
forward-backward asymmetry in Drell-Yan production for the SM (solid curve)
and with 200 GeV scalar leptoquark exchange, assuming $\lamt=1$, with
left-(right-)handed couplings to u-quarks, corresponding to the dashed
(dotted) curve, and left-(right-)handed couplings to d-quarks (dash-dotted
curve for both cases).}
\label{dylqs}
\end{figure}

The invariant mass distribution is obtained by integrating the
differential cross section over the regions $-Y\leq y\leq Y$ and
$-Z\leq z\leq Z$ where
\begin{eqnarray}
Y & = & min[y_{max}, -ln(M/\sqrt s)] \,,\\
Z & = & min[\tanh(Y-|y|),1] \nonumber
\end{eqnarray}
where $y_{max}$ represents the rapidity coverage of the detector or of the
applied cuts.  To calculate
the forward-backward asymmetry, Eq. \ref{dyan} is first integrated over
the forward ($z>0$) and backward ($z<0$) regions separately (subject to
$|z|\leq|Z|$) and then over $y$; the difference of the forward and backward
cross sections divided by their sum then gives $A_{FB}$.  Explicitly, we
define
\begin{eqnarray}
{d\sigma^+\over dM} & = & \left[ \int_{y>0}dy+\int_{y<0}dy\right]
\left[ \int_{z>0}+\int_{z<0}\right] \left[ {d\sigma\over dMdydz}\right] dz \,,\\
{d\sigma^-\over dM} & = & \left[ \int_{y>0}dy\pm\int_{y<0}dy\right]
\left[ \int_{z>0}-\int_{z<0}\right] \left[ {d\sigma\over dMdydz}\right] dz 
\,,\nonumber
\end{eqnarray}
where the $+\,, (-)$ sign is relevant for $\bar pp\,, (pp)$ collisions.  The
forward-backward asymmetry is then given by
\begin{equation}
A_{FB}(M) = { {d\sigma^-/ dM}\over {d\sigma^+/ dM} }\,.
\end{equation}
Figure \ref{dylqs} displays our results for (a) the Drell-Yan invariant mass 
spectrum and (b) forward-backward asymmetry in the electron channel with 
and without scalar leptoquark exchange at the Tevatron.  Here, we have 
employed the present rapidity coverage of the CDF detector as used in their
Drell-Yan analysis\cite{cdfdy} $|y_{max}|\leq 1$.  We have assumed a scalar
leptoquark mass of 200 GeV and Yukawa coupling strength of $\lambda/e=1$.
In this figure, the SM is represented by the solid curve, and the cases with
left-, right-handed leptoquark couplings to u-(d)quarks correspond to the
dashed(dot-dashed), dotted(dot-dashed) curves.  We see that the influence of
leptoquark exchange on this process in minimal, even for these large values
of the couplings!  It is clear that at the present level of statistics, the
Tevatron experiments are not sensitive to leptoquark exchange in Drell-Yan
production.

\vspace*{-0.5cm}
\nn

\begin{figure}[htbp]
\centerline{
\psfig{figure=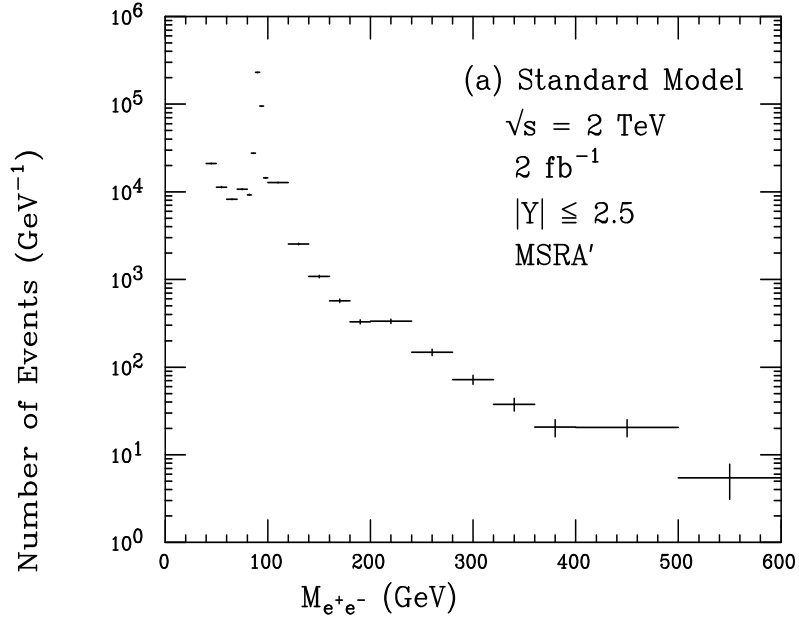,height=10.cm,width=12cm,angle=-90}}
\vspace*{-5mm}
\centerline{
\psfig{figure=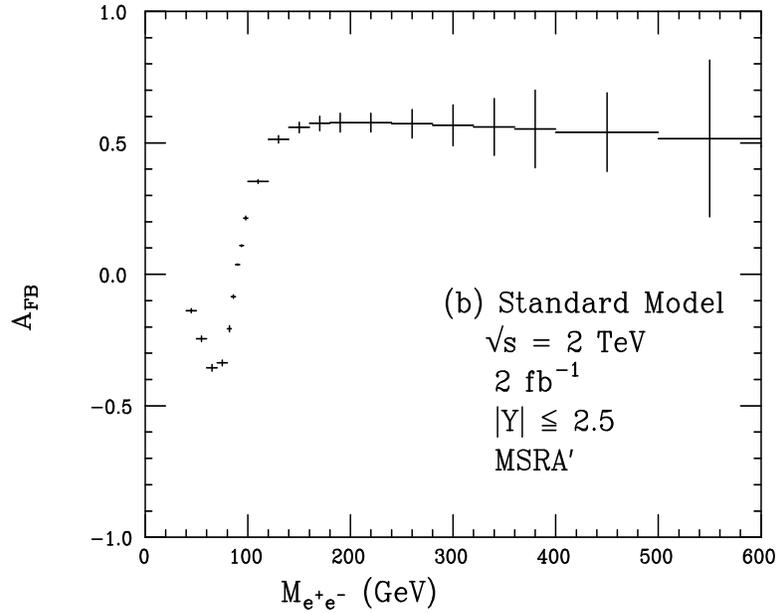,height=10.cm,width=12cm,angle=-90}}
\vspace*{-0.75cm}
\caption{Bin integrated lepton pair (a) invariant mass distribution and
(b) forward-backward asymmetry for Drell-Yan production in the SM at the
Main Injector.  The vertical lines correspond to the expected statistical
error in each bin.}
\label{binint}
\end{figure}

\begin{figure}[htbp]
\centerline{
\psfig{figure=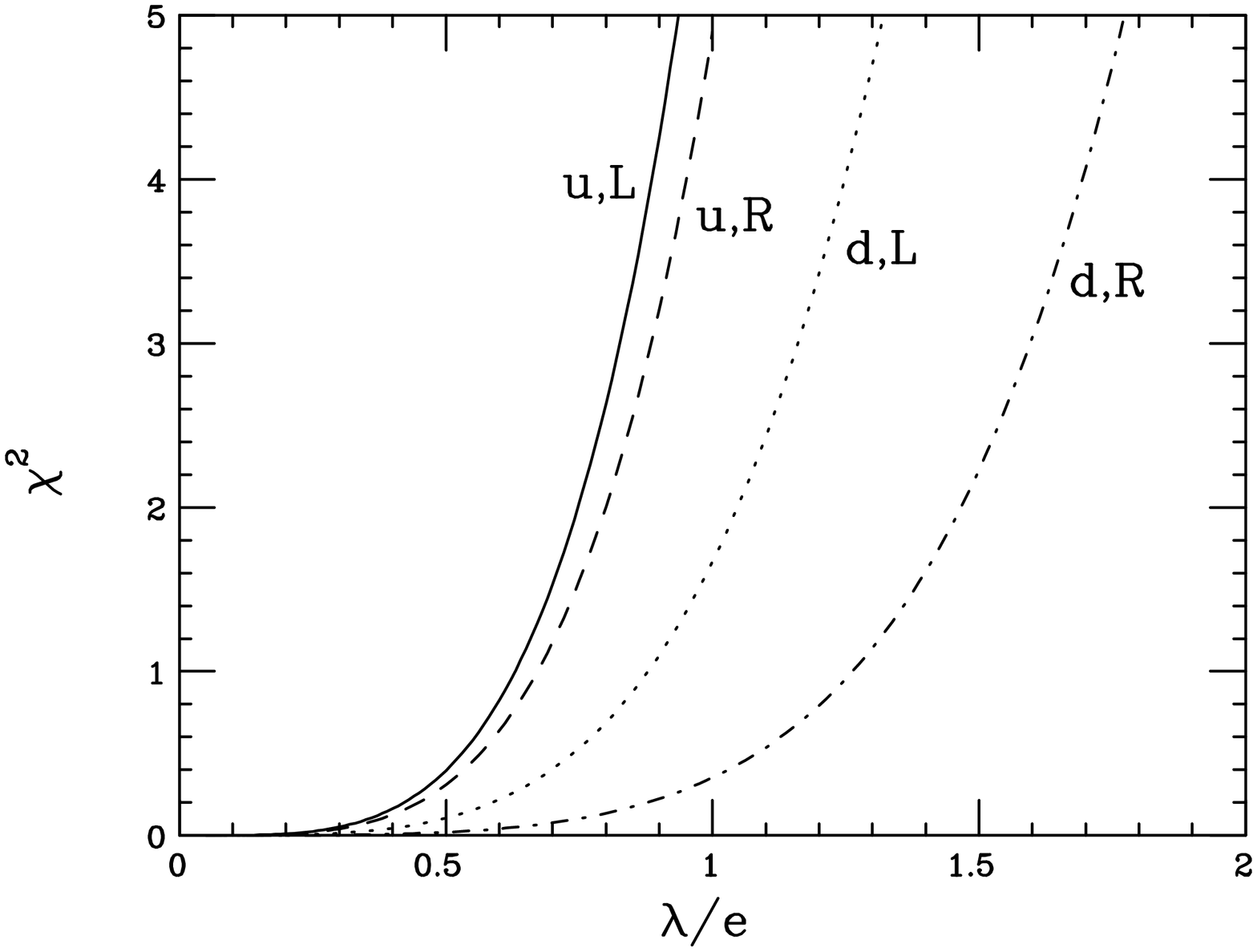,height=14.cm,width=16cm,angle=-90}}
\vspace*{-0.75cm}
\caption{$\chi^2$ fits to Drell-Yan production including the effects of a
200 GeV scalar leptoquark for each type of leptoquark coupling as labeled.
The $95\%$ C.L. constraints are obtained when $\chi^2=3.842$.}
\label{chi2dy}
\end{figure}

We next examine the level of sensitivity that will be achievable at the
Main Injector with $2$ fb$^{-1}$ and $\sqrt s=2\tev$.  Following 
Ref. \cite{tevstudy}, we enlarge the rapidity coverage to $|Y|<2.5$ and
construct 21 invariant mass bins, corresponding to
\begin{eqnarray}
{\mbox{4 bins in steps of 10 GeV in the range}} & 40\leq M\leq 80 & 
{\mbox{GeV}}\,,\nonumber\\
{\mbox{5 bins in steps of 4 GeV in the range}} & 80\leq M\leq 100 &
{\mbox{GeV}}\,,\nonumber\\
{\mbox{5 bins in steps of 20 GeV in the range}} & 100\leq M\leq 200 &
{\mbox{GeV}}\,,\\
{\mbox{5 bins in steps of 40 GeV in the range}} & 200\leq M\leq 400 &
{\mbox{GeV}}\,,\nonumber\\ 
{\mbox{2 bins in steps of 100 GeV in the range}} & 400\leq M\leq 600 &
{\mbox{GeV}}\,.\nonumber
\end{eqnarray}
The bin integrated cross section and asymmetry are then obtained for both the 
SM and for the case with 200 GeV scalar leptoquark exchange.  We determine
the statistical error on these quantities in each bin, which are taken to be
$\delta N=\sqrt N$ and $\delta A=\sqrt{(1-A^2)/N}$.  The bin integrated
results for the SM, along with the error associated with each bin,
are displayed in Fig. \ref{binint}. To evaluate what
constraints may be placed on the leptoquark  coupling we then perform a
$\chi^2$ analysis according to the usual prescription,
\begin{equation}
\chi^2_i=\sum_{\rm bins} \left( {Q_i - Q_i^{SM}\over\delta Q_i}
\right)^2 \,,
\end{equation}
where $Q_i$ represents each observable quantity.  The resulting $\chi^2$ 
distribution, summing over both observables, is presented in Fig. \ref{chi2dy}
as a function of $\lambda/e$ for the various scenarios of left-, right-handed
leptoquark couplings to u- and d-quarks.  We see that the $95 \%$ C.L. bounds
(corresponding to $\Delta\chi^2=3.842$) 
on \lamt, are quite weak and are inferior
to the present restrictions from low-energy data.

\vspace*{-0.5cm}
\nn

\vspace*{-0.5cm}
\nn

We briefly summarize this section by pointing out (i) present Tevatron data
analyses easily
allow for scalar leptoquarks in the 200-210 mass range, but exclude
vector leptoquarks and leptogluons.  (ii) If 200 GeV leptoquarks exist and 
are responsible for the event excess observed at HERA, the Main
Injector will observe them in both the pair and single production mode, and
can confirm the values of the mass and coupling observed at HERA.  (iii) 
However,
since these production mechanisms are QCD processes, hadron colliders can
not provide any information on the electroweak properties of leptoquarks, or
on the chiralities of their couplings.  The same conclusions hold for the LHC.

\section{Constraints from Precision Electroweak Measurements}

\begin{figure}[htbp]
\centerline{
\psfig{figure=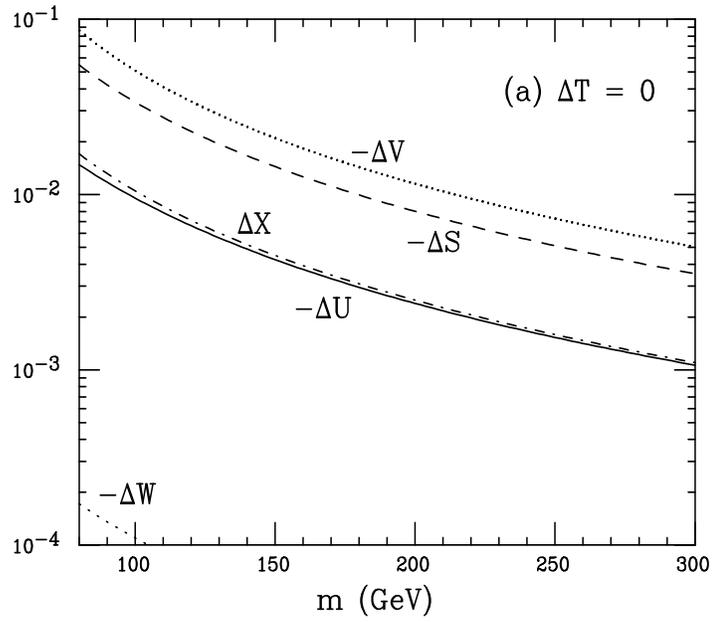,height=10cm,width=12cm,angle=-90}}
\vspace*{-.75cm}
\centerline{
\psfig{figure=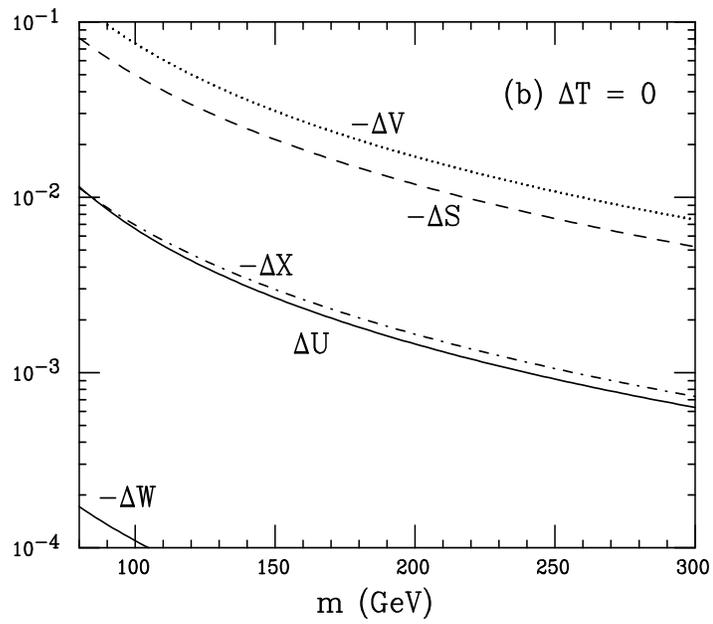,height=10cm,width=12cm,angle=-90}}
\vspace*{-0.6cm}
\caption{Shifts in the oblique parameters as functions of the leptoquark mass 
for the (a) $\widetilde R_{2L}$  and (b) $R_{2L}$ or $R_{2R}$ cases.}
\label{stuvwx}
\end{figure}

With masses of only $\sim 200$ GeV, which is not far above the top quark mass,
one may wonder if such light leptoquarks have any influence on $Z$-pole 
precision measurements.  We first examine the oblique 
parameters{\cite {prec,obl}}, following the conventions 
of Peskin and Takeuchi, and 
in particular the extension by Maksymyk, Burgess and London to the enlarged
parameter set S,T,U,V,W and X. The additional parameters V, W, and X 
need to be included
when new light particles (with masses of order the electroweak scale or less)
are introduced.  Since we are assuming that the various 
leptoquark multiplets are degenerate, apart from higher order corrections,
they do not contribute to the parameter T at one-loop (we recall that T 
is a measure of the
mass splitting between the particles in a weak isospin multiplet).  However,
we do expect finite one-loop corrections to the other 
parameters, which can be straightforwardly calculated for each 
of the leptoquark multiplets.  Here we present the results 
for the case of the $F=0$ type leptoquarks, $\widetilde R_{2L}$ and
$R_{2L,R}$, in Fig.~\ref{stuvwx}, which displays
the shifts in the remaining oblique parameters as a function of the leptoquark
mass.  The results for the other leptoquark
cases are found to be quite similar. As can be easily 
seen from the figure, leptoquarks in this mass range do not make appreciable 
contributions to the oblique parameters. 

\vspace*{-0.5cm}
\nn
\vspace*{0.4mm}

In addition to oblique corrections, it is possible that relatively 
light leptoquarks can lead to substantial vertex corrections, \eg, in the case 
of $Z\to e^+e^-$, where first generation leptoquarks and quarks may 
contribute in a loop.  This case has been previously examined by several 
authors{\cite {vertex}}. However, as shown by both Eboli \etal\
and Bhattacharyya \etal,
the fact that first generation leptoquarks couple only to $u$-  or $d$-quarks 
leads to a substantial suppression of their potential contribution 
to this vertex. For leptoquark masses of order 200 GeV, Yukawa couplings
of order unity cannot be excluded by these considerations.  We see that these
constraints are much weaker than those imposed from low-energy data.

\section{$e^+e^-$ Colliders}

\begin{figure}[htbp]
\centerline{
\psfig{figure=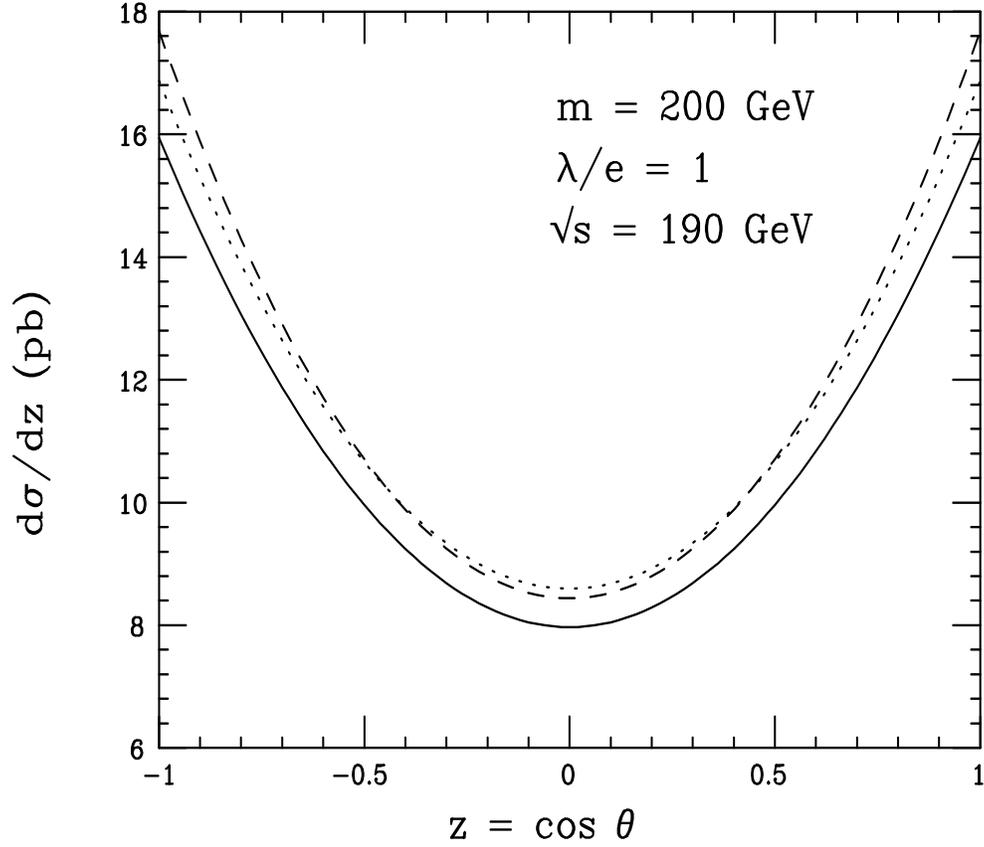,height=14cm,width=16cm,angle=-90}}
\vspace*{-0.9cm}
\caption{Symmetrized angular distribution of $q\bar q$ final states in 
$e^+e^-$ annihilation at 190 GeV with $z=\cos \theta$. The solid 
curve is for the SM  while the dotted(dashed) curve includes the 
contribution of a 200 GeV 
scalar(vector) leptoquark coupling to $d_L$ with $\tilde \lambda=1$.}
\label{eeqqplot}
\end{figure}

  There are several ways in which leptoquarks may make their presence known 
in $e^+e^-$ collisions\cite{us,more,others}.  
At center of mass energies below the threshold for
pair production, the existence of leptoquarks can lead to 
deviations{\cite {us,others}} in the cross section and angular distributions
for $e^+e^- \to q\bar q$. This may be particularly 
relevant when $\sqrt s$ is comparable to the leptoquark mass as would be the 
case at LEP II if a 200 GeV leptoquark did exist. The origin of 
these modifications is due to the $t-$($u-$)channel exchange of the $F=0(2)$ 
leptoquark and is thus proportional in amplitude to the square of the 
unknown Yukawa coupling. Including such terms results in the 
$e^+e^- \to q\bar q$ tree-level differential cross section given 
in the Appendix.
Note that for first generation leptoquarks either the 
$u\bar u$ or/and $d\bar d$ final state may be influenced, depending on the 
leptoquark species being exchanged.  There will be no 
effect on $s\bar s$, $c\bar c$ or $b\bar b$ final states.  We now examine the
sensitivity of the cross section to the value of the scaled coupling
$\tilde \lambda$.   As an example, we consider the case of a 
200 GeV leptoquark coupling to $d_L$ with $\tilde \lambda_L=1$ at LEP II with 
$\sqrt s=190$ GeV. Since d-quarks cannot easily be distinguished from any of 
the other light flavors, nor can quarks be differentiated from antiquarks 
without some difficulty{\cite {qfb}}, we have symmetrized the expression in
the Appendix with respect to $\cos \theta$ and summed over the possible
light quark final states. Owing to this symmetrization, 
independent sensitivity to new $t-$ versus $u-$ channel exchange is lost
and we can no longer 
distinguish $F=0$ from $F=2$ type leptoquarks. Recall that in this sample
case only $d_L$ is 
assumed to couple to the leptoquark, hence this flavor summation will
significantly degrade the sensitivity to the Yukawa couplings in this process.
Figure~\ref{eeqqplot} displays the resulting angular distributions for the SM
and scalar (and for completeness, vector) leptoquarks. From this it is clear 
that even large values of the Yukawa coupling do not lead 
to sizeable changes in the shape and size of the 
cross section. Of course there is nothing special about our sample case of
leptoquarks coupling 
to $d_L$ and we expect modifications of a similar size when the 
leptoquark couples instead to $d_R$ or $u_{L,R}$. 

\vspace*{-0.5cm}
\nn
\vspace*{0.4mm}

\begin{figure}[htbp]
\centerline{
\psfig{figure=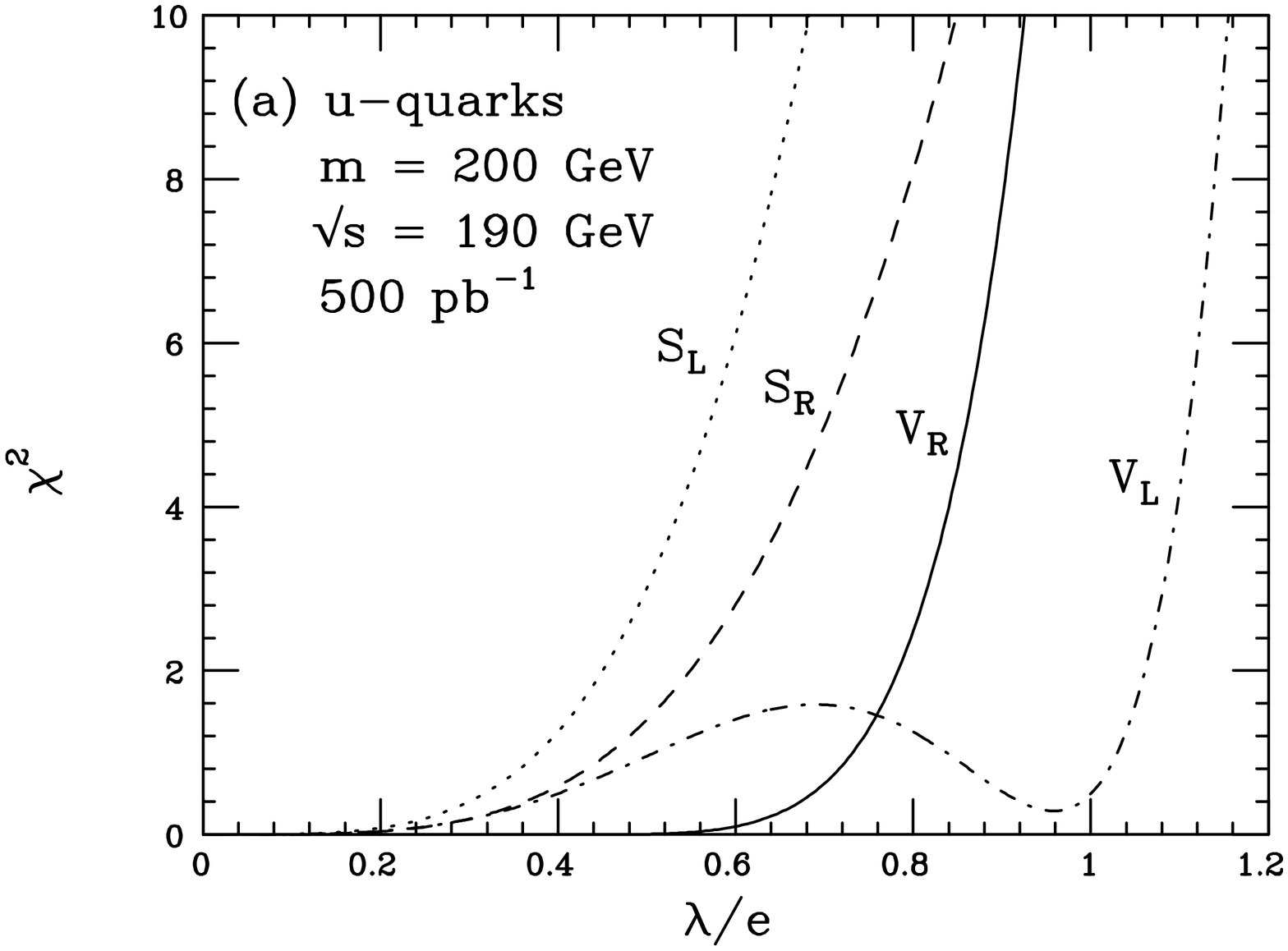,height=10cm,width=12cm,angle=-90}}
\vspace*{-.75cm}
\centerline{
\psfig{figure=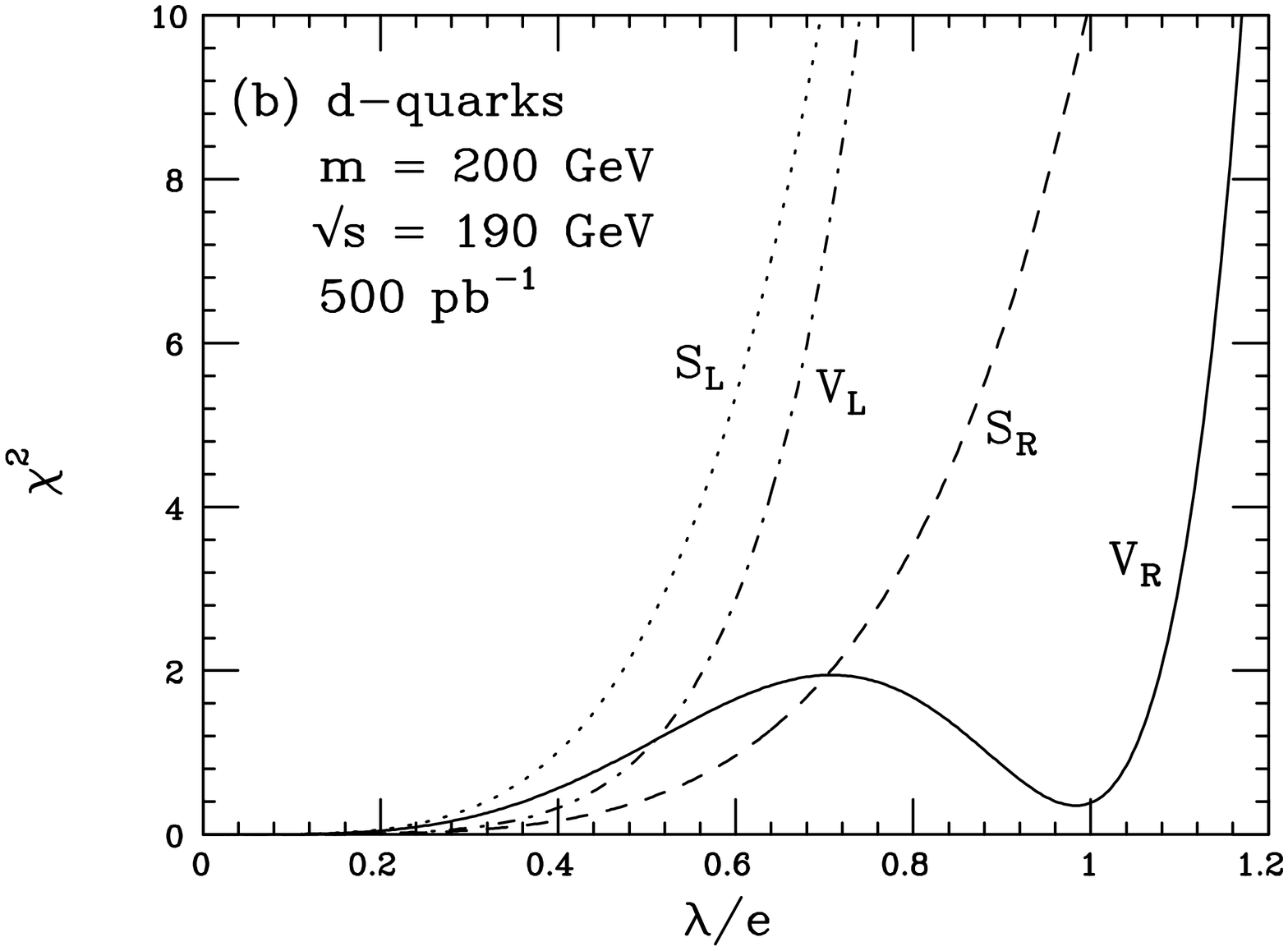,height=10cm,width=12cm,angle=-90}}
\vspace*{-0.6cm}
\caption{$\chi^2$ fits to the SM angular distribution for $e^+e^-\to q\bar q$ 
at 190 GeV including the effects of a 200 GeV leptoquark coupling to (a) u-
or (b) d-quarks. In both cases the dotted(dashed) curve corresponds to a scalar 
leptoquark with a left(right)-handed coupling while the dash-dotted(solid) 
curve corresponds to the vector leptoquark case with left(right)-handed 
couplings. The $95\%$ CL limits are obtained when $\chi^2=3.842$.}
\label{chi2l}
\end{figure}

We now estimate the potential sensitivity of LEP II cross section measurements 
to non-zero values of the Yukawa coupling by generating a Monte Carlo 
data sample of $q\bar q$ events 
at 190 GeV with an integrated luminosity of $500\inpb$.  The cross section 
is divided into 10 $\cos \theta$ bins of identical size and no additional 
cuts are applied. We then determine the sensitivity to the size of the Yukawa 
couplings assuming that only one of the quark final states couples 
to the leptoquark by performing a $\chi^2$ procedure.  The results of this 
analysis are shown in Fig.~\ref{chi2l} for leptoquarks of either spin and for 
both coupling helicities to up and down quarks. 
For scalar leptoquarks we expect that the LEP II limits on $\tilde \lambda$ 
will lie in the 0.5-0.8 range which is somewhat less restrictive than 
those that already exist due to the APV data and the $\pi \to e\nu$ decay. 
The expected bounds 
are also about a factor of 5 larger than the typical values necessary to 
explain the HERA excess in terms of leptoquarks.  The cases where the 
leptoquarks are in a multiplet and can couple to {\it both} $u$ and 
$d$ quarks do not lead to any substantial improvement in these constraints when 
these contributions are combined. It thus 
appears that LEP II will be insensitive to any leptoquark consistent with the 
HERA data.

\vspace*{-0.5cm}
\nn
\vspace*{0.1mm}

  At higher energy $e^+e^-$ colliders such as the NLC, leptoquark pairs can be 
produced directly and their properties examined{\cite {us}} in detail.
As we will see, this allows us to easily identify which leptoquark is being 
produced.  In what follows we again
limit our discussion to the spin-0 case since it is directly 
relevant for the HERA events.
The pair production differential cross section is given by
\begin{eqnarray}
{d\sigma\over dz} & = & {3\pi\alpha^2\over 8s}\beta^3(1-z^2)\left\{ 
2\sum_{ij}(v^i_ev^j_e+a^i_ea^j_e)C_iC_jP_{ij} \right. \\
& &\left. -{s\over t}\sum_i\left[ \lamt^2_L(v^i_e+a^i_e)+\lamt^2_R(v^i_e-
a^i_e)\right] C_iP_i +{s^2\over 4t^2}(\lamt^4_L+\lamt^4_R)\right\} \,,\nonumber
\end{eqnarray}
where the sum extends over $\gamma$ and $Z$ exchange,
$\beta=\sqrt{1-4m^2/s}$, $C_1=Q_{LQ}$, $C_2=2[T_3-x_wQ]_{LQ}(\sqrt{2}G_F
M_Z^2/4\pi\alpha)^{1/2}$, and $v\,, a\,, P_{ij}\,,$ and $P_i$ are defined in
the Appendix.

When large Yukawa couplings are present, the exchange of $u$ or $d$ quarks in 
the $t-$channel can seriously modify the pair production angular distribution 
away from the conventional $\sin^2\theta$ dependence leading to an appreciable 
forward-backward asymmetry. However, for $\tilde \lambda \leq 0.1$ this 
asymmetry is found to be below the $0.2\%$ level at a $\sqrt s=500$ GeV 
collider when 200 GeV leptoquarks are being produced.  Thus, in the limit
that $\lamt\approx 0$, the cross 
section for a particular final state, such as $\ell \ell jj$, depends solely 
on the electroweak quantum numbers of the members of the relevant leptoquark 
multiplet and their branching fractions to charged leptons for a fixed
leptoquark mass.   Further
information on the leptoquark electroweak quantum numbers may be obtained from
examining the left-right polarization asymmetry, defined as usual as
\begin{equation}
A^{LR}_{\ell\ell jj} \equiv {\sigma_L(\ell\ell jj)-\sigma_R(\ell\ell jj)\over
\sigma_L(\ell\ell jj)+\sigma_R(\ell\ell jj)}\,.
\end{equation}
Table~5 
summarizes the production cross sections for each final state as well as the 
polarization asymmetry associated with the $\ell \ell jj$ final state for all 
scalar leptoquark multiplets.
 It is clear from the Table that by measuring 
the rates for each final state channel in addition to the polarization 
asymmetry that the 
identity of the produced leptoquark would be straightforward to obtain 
assuming the design luminosity of $50 fb^{-1}$. We note in passing that since 
the Yukawa couplings of the $\sim 200$ GeV LQ's are so small it is possible 
that LQ pair bound states(LQ-onium?) can form in the mass region near 400 
GeV. These states can be produced in a number of ways, such as $WW$-fusion, 
and can only be explored in detail at a lepton collider such as the NLC 
although the lowest lying states can be produced at a hadron collider through 
gluon fusion. 

\begin{table*}[htbp]
\leavevmode
\begin{center}
\label{lqnlc}
\begin{tabular}{lcccc}
\hline
\hline
Leptoquark&$\ell \ell jj$&$\ell\nu jj$&$\nu\nu jj$&$A^{LR}_{\ell \ell jj}$ \\
\hline

$S_{1L}$        & 1.88 & 3.77 & 1.88 & -0.618 \\
$S_{1R}$        & 7.53 & 0.0  & 0.0  & -0.618 \\
$\widetilde S_{1R}$ &120.4 & 0.0  & 0.0  & -0.618 \\
$S_{3L}$        &192.2 & 3.77 & 1.88 &  0.931 \\
$R_{2L}$        &181.0 & 0.0  & 80.4 &  0.196 \\
$R_{2R}$        &261.4 & 0.0  & 0.0  & -0.141 \\
$\widetilde R_{2L}$ & 47.6 & 0.0  & 33.2 &  0.946 \\
\hline
\hline
\end{tabular}
\caption{ Cross sections for the three leptoquark pair decay channels in fb at 
a 500 GeV NLC assuming complete leptoquark multiplets with a common mass of 
200 GeV. The polarization asymmetry in the $\ell \ell jj$ channel is also 
given. In all cases $\tilde \lambda \ll 1$ is assumed.}
\end{center}
\end{table*}

Single leptoquark production at the NLC is also possible via e$\gamma$ 
collisions\cite{egam}, with a production rate which is quite sensitive to
the electric charge of the leptoquark.  The amplitude for this
process is proportional to the Yukawa coupling $\lambda$  and results in a
cross section which is not significantly different in magnitude from
that of pair
production if $\tilde \lambda$ is not far from unity.  If both electron and 
photon beam polarization is available, asymmetries  can also be used to
determine the leptoquark's quantum numbers as has been demonstrated by 
Doncheski and Godfrey{\cite {egam}}. Table~6 shows the 
production rate for a 200 GeV leptoquark in $\gamma e$ collisions at the NLC 
for a luminosity of $50fb^{-1}$ using either the backscattered laser or 
Weisacker-Williams photon spectra. In obtaining these results, the hadronic 
content of the photon has been ignored; its inclusion would somewhat increase 
these rates. The backscattered laser approach has the advantage of a harder 
spectrum (although it cuts off at $x \simeq 0.84$) and both beams can be 
polarized. In the Weisacker-Williams case an additional factor of 2 is 
included since both $\gamma e^{\pm}$ collisions are possible. 
Even for this small value of $\tilde \lambda$ the production
rates are at an observable level.  
It is clear that by using these rates together 
with the use of beam polarization the quantum numbers can be determined 
in a straightforward manner.

\begin{table*}[htbp]
\leavevmode
\begin{center}
\label{lqs}
\begin{tabular}{lccc}
\hline
\hline
Leptoquark& Backscattered Laser & Weisacker-Williams & $B_{\ell}$ \\
\hline

$S_{1L}$        &      212.  &  56.8  & 0.5 \\
$S_{1R}$        &      212.  &  56.8  & 1   \\
$\widetilde S_{1R}$ &  109.  &  24.4  & 1   \\
$S_{3L}$        &      430.  &  106.  &$\simeq 0.75$ \\
$R_{2L}$        &      332.  &  79.5  & 1\\
$R_{2R}$        &      381.  &  92.6  & 1\\
$\widetilde R_{2L}$ &  49.1  &  13.1  & 1\\
\hline
\hline
\end{tabular}
\caption{Rates for single leptoquark production in $\gamma e$ collisions at 
a 500 GeV NLC assuming complete leptoquark multiplets with a common mass of 
200 GeV and an integrated luminosity of $50fb^{-1}$. In all cases 
$\tilde \lambda =0.1$ is assumed and a $p_T$ cut on the quark jet of 10 GeV 
has been applied. The charged lepton branching fraction for the produced 
multiplet is also given.}
\end{center}
\end{table*}

\section{Unification with Leptoquarks}

At this point one may wonder how scalar leptoquarks of the $F=0$ type would 
fit into a larger picture. As we saw earlier, both $R_{2L,R}$ can be embedded 
into a {\bf 45}, $\overline{\mbox{\bf 45}}$ representation of $SU(5)$ while 
$\widetilde R_{2L}$, which has less exotic 
electric charges, can be placed in a {\bf 10} or {\bf 15}. In a SUSY 
extension, where one normally adds complete multiplets to automatically 
insure coupling constant unification, we 
would thus need to add either a {\bf 10}$+\overline{\mbox{\bf 10}}$
({\bf 15}$+\overline{\mbox{\bf 15}}$) or a {\bf 45}$+\overline{\mbox{\bf 45}}$
at low energies. Here the barred representation is 
introduced to avoid anomalies and 
to guarantee that the fermionic components are vector-like with respect to the 
SM gauge group. As is well-known, the addition of extra matter representations 
delays unification and brings the GUT scale much closer to the string scale. 
A short analysis shows that adding 
complete {\bf 15}$+\overline{\mbox{\bf 15}}$'s or 
{\bf 45}$+\overline{\mbox{\bf 45}}$'s would lead to a dramatic loss of 
asymptotic freedom(AF) at 
one loop (\ie, $\beta_i>0$) and, in the later case, both $R_{2L,R}$ would be 
present in the low 
energy spectrum. We would then need to explain why only one of the 
chiral couplings was present as well as the generational structure of the 
couplings by the imposition of some extra symmetries. 
The addition of the 
{\bf 10}$+\overline{\mbox{\bf 10}}$ to the usual MSSM particle content does 
not lead to either the loss of AF at one loop or to the 
problem of suppressing one of the chiral couplings. Interestingly, these 
general considerations tell us that the {\it only} leptoquarks consistent 
with both SUSY and unification within standard 
$SU(5)$ are $\widetilde R_{2L}$ and $S_{1L,R}$, the 
later being the familiar leptoquarks of $E_6$ string-inspired 
models{\cite {phyrep}}. In both cases the QCD beta-function 
is found to vanish at one loop (in the $E_6$ case three 
{\bf 5}$+\overline{\mbox{\bf 5}}$'s are 
present with a different leptoquark for each generation). Since we can safely 
add only a single {\bf 10}$+\overline{\mbox{\bf 10}}$ at low energies, 
a realistic model would still need to 
explain the hierarchy of generation dependent coupling strengths. 
We note that in the light {\bf 10} together 
with the leptoquark will be a $Q=-1$ isosinglet bilepton{\cite {dc}} and a 
$Q=2/3$, color triplet, isosinglet diquark. 

Although the leptoquarks in the {\bf 10}$+\overline{\mbox{\bf 10}}$ are 
found to be consistent with both unification and AF considerations, we still 
cannot identify them with the source of the HERA events due to the nature of 
their $SU(5)$ coupling structure. To form a $SU(5)$ singlet in the product 
$\bar 5_i\bar 5_j 10_{k}$, only the antisymmetric terms in the $i,j$ can 
contribute. This would imply that the leptoquark must couple in an 
antisymmetric fashion with respect to the generations{\cite {worah}} so that 
the phenomenologically required $e^+d$-type coupling would be prohibited. 
Thus the requirements of AF, SUSY $SU(5)$ unification, and the addition of 
complete multiplets do not simultaneously allow any $F=0$ leptoquarks with 
couplings to only a single generation.

How do we circumvent this result? One possibility is to surrender the 
assumption of the addition of complete $SU(5)$ representations at low 
energies. This certainly allows us more flexibility at the price of 
naturalness but still requires us to chose subsets of 
$SU(3)_C\times SU(2)_L \times U(1)_Y$ representations from the {\bf 10}, 
{\bf 15} or {\bf 45} which maintain AF and unification. Except for the rather 
bizarre choice of adding a $(2,3)(1/6)$ from {\bf 15} and a
$(1,1)(1)\oplus (1,\bar 3)(-2/3)$ from a {\bf 10} at low energy, a short 
analysis shows that no other 
solutions were found to exist. Here the notation refers to the 
$(SU(3)_C,SU(2)_L)(Y/2)$ quantum numbers of the representation. Thus apart 
from this exotic choice we find that our constraints are sufficiently strong 
as to disallow any $F=0$ leptoquarks at low energy in the SUSY $SU(5)$ 
context.

It is clear that we must give up conventional 
$SU(5)$ if we want a HERA-inspired leptoquark 
in a SUSY-GUT framework{\cite {king,string}}. Perhaps the most attractive 
scenario for this is the 
flipped $SU(5)\times U(1)_X$ model{\cite {nan}} wherein the SM fermion 
content is extended by the addition of the 
right-handed neutrino, $\nu^c$, and the 
conventional roles of $u^c$ and $e^c$ are interchanged with those of $d^c$ 
and $\nu^c$. Thus, $u^c$ lies in the $\overline{\mbox{\bf 5}}$,  
the $d^c$ lies in the {\bf 10} and $e^c$ is in an $SU(5)$ singlet. In this 
case, completely different, and successful from the HERA point of view, 
leptoquark embeddings are now possible. 
For example, $R_{2R}$ can be placed in a {\bf 10}$+\overline{\mbox{\bf 10}}$ 
without the difficulties associated with the cross generational 
couplings we encountered above since $e^c$ is an $SU(5)$ singlet. In this 
case we would still need to impose 
some additional symmetries so that $R_{2R}$ could only couple to the first 
generation. Note that this {\bf 10}$+\overline{\mbox{\bf 10}}$ would also 
contain the $\widetilde S_{1R}$ leptoquark as well as an isosinglet bilepton 
with $Q=1$. The $\widetilde S_{1R}$ may also show up as a separate resonance 
in $e^-p$ collisions as discussed above if it has Yukawa couplings of order
$\tilde \lambda \sim 0.1$ once sufficient luminosity is accumulated.  
$R_{2L}$ can lie in a {\bf 10}$+\overline{\mbox{\bf 10}}$, but 
would require cross generational coupling as above since both $L$ and $u^c$ 
are in the $\overline{\mbox{\bf 5}}$. On the otherhand, $\widetilde R_{2L}$ now 
lies in a {\bf 45}$+\overline{\mbox{\bf 45}}$ and is excluded by the AF 
constraints.  It thus appears that the flipped $SU(5)\times U(1)_X$ scenario 
provides a natural embedding for at least one of these $F=0$ leptoquarks,  
$R_{2R}$, and may predict the simultaneous existence of an $F=2$ leptoquark. 
Other GUT groups may provide phenomenologically successful 
embeddings for the other $F=0$ leptoquarks.
The extension of the spectrum to include the 
$\nu^c$ field may introduce some new additional interesting phenomenological 
implications for these leptoquarks since their interactions now extend 
beyond those described by the Lagrangian in Section 2 and may possibly yield 
excess events in the charged current channel. 

If LQ's really exist in a SUSY framework then their fermionic partners, 
leptoquarkinos(LQ-inos) must also be observed{\cite {lqi}}. What are the 
masses of such states? If the LQ-ino is lighter than the LQ then the partner 
will decay directly to the LQ plus the LSP, since the LQ Yukawa is so small,  
unless the two states are nearly degenerate. This implies that the LQ-ino is 
the more massive of the two states, thereby decaying into the LQ plus LSP. 
Since the fermion pair cross section is larger than the scalar pair cross 
section at the Tevatron when both are color triplet states, it is clear that 
the mass of the LQ-ino must be substantially heavier than 200 GeV in order to 
reduce the anticipated number of $eejj+\slash p_T$ events to an acceptable 
level.

\begin{figure}[htbp]
\centerline{
\psfig{figure=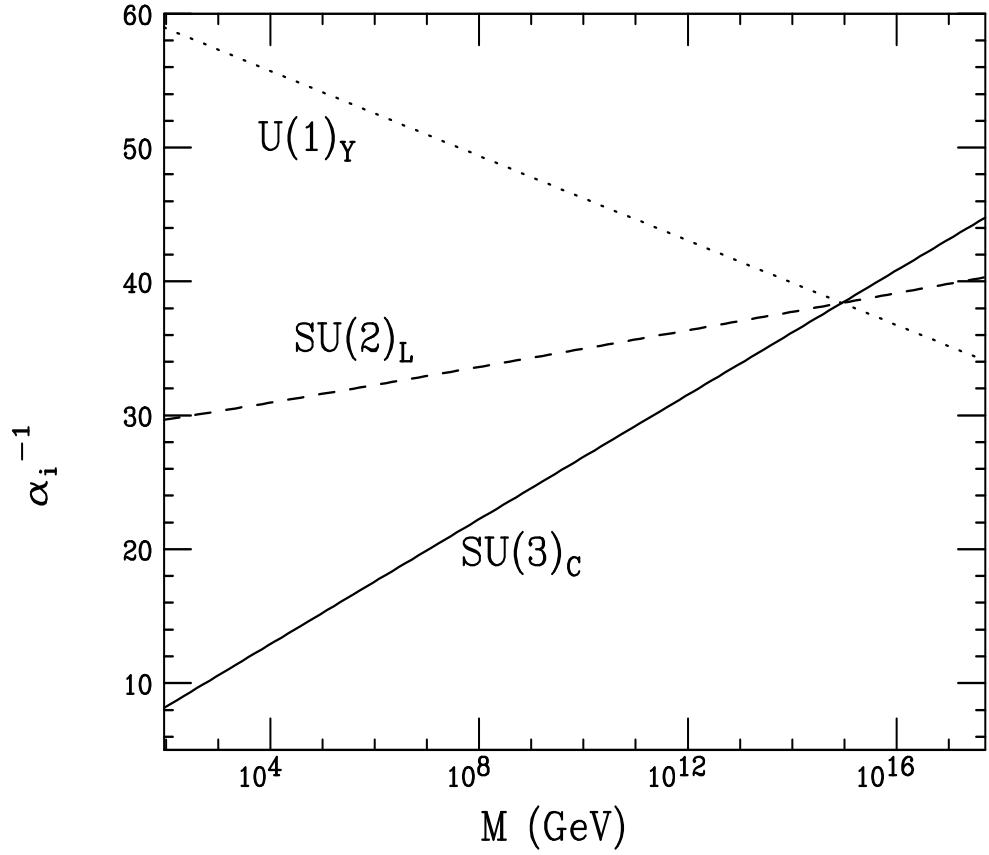,height=14cm,width=16cm,angle=-90}}
\vspace*{-0.9cm}
\caption{Two-loop RGE evolution of the model with the SM particle content 
together with a pair of $\widetilde R_{2L}$ type leptoquarks and an additional 
Higgs doublet.}
\label{guts}
\end{figure}

What can we say about these leptoquarks in the non-SUSY context? Here we can 
only be more speculative. 
As is well-known, unification attempts without SUSY using only the SM particle 
content are doomed to failure in that they predict a too small value of the 
unification scale, implying a rapidly decaying proton, and lead to values of 
$\alpha_s(M_Z)$ which are smaller than the experimentally determined value 
by many standard deviations. One is led to 
consider the general question of whether one could add ad hoc sets of 
additional (non-SUSY) particles with masses at the electroweak scale to those 
which already exist within the SM to get 
unification at a higher scale and a proper value of $\alpha_s(M_Z)$ by 
sufficiently modifying the SM beta functions. A short consideration shows that 
this is difficult to arrange. At one-loop, the modified beta functions must 
satisfy the so-called ``B-test''{\cite {mp}}: 
\begin{equation}
B = {b_3-b_2\over {b_2-b_1}} = 0.719\pm 0.01 \pm 0.04   \,, 
\end{equation}
where $B_{SUSY}=5/7=0.714$ clearly satisfies the test. In earlier 
work{\cite {oldt}}, many additional particles with a wide range of strong and 
electroweak quantum numbers were added in many thousands of combinations 
in order to attempt to
satisfy these constraints with only two dozen candidates surviving (see Table 1 
in Ref.{\cite {oldt}}). Given the higher precision of current data, at least 
several of these survivors could now be eliminated leaving  a very short 
list. A survey of this list shows that there is 
only one case with scalars which have the 
correct quantum numbers to be consistent with leptoquarks of {\it any} kind. 
Interestingly, 
this case corresponds to a pair of $\widetilde R_{2L}$ 
leptoquarks with the Higgs 
sector of the SM augmented by an additional doublet. This scenario was first
discovered in the analysis of Murayama and Yanagida{\cite {my}} and is quite 
unique with $B=0.693$. 
Figure~\ref{guts} shows a two-loop renormalization group(RGE) analysis of this 
particular case. 
Interestingly, for $\sin^2\theta_w(\overline {MS})=0.23165$ and 
$\alpha_{em}^{-1}=127.90$ we obtain 
$\alpha_s(M_Z)=0.123$ and a proton lifetime{\cite {mrbill}} of $10^{32\pm 1}$ 
yrs, which is close to the present limit in the $e^+\pi^0$ mode{\cite {pdg}}. 

\vspace*{-0.5cm}
\nn
\vspace*{0.4mm}

As a final comment we note the oft-neglected problem of R-parity violation 
within a GUT context. A term in the superpotential of the form 
$\lambda_{ijk}\bar 5_i\bar 5_j 10_{k}$ would generate all of the usual 
lepton and baryon number violating R-parity violating terms 
{\it simultaneously} with comparable Yukawa couplings and would lead to 
a very rapid proton decay. It may be possible, however, to forbid such an 
R-parity violating term at the renormalizable level while having them arise as 
higher dimensional operators{\cite {tamv}}. It is not clear, however, that 
an interaction of the type $e^cd\tilde u$ can be generated in this approach 
with a Yukawa coupling of the right magnitude to explain the HERA 
events{\cite {giu}}.

\section{Conclusions}

In this paper we have considered the detailed phenomenological implications 
of interpreting the excess of events observed by both H1 and ZEUS at HERA as 
the production of an $s$-channel scalar leptoquark resonance. First, 
we demonstrated that the D0 leptoquark search data strongly indicate that 
this resonance could be neither a leptogluon nor a vector leptoquark with a 
mass near 200 GeV due to their much larger pair production cross sections at
the Tevatron. Secondly, we 
showed that the HERA data 
itself, in particular the apparent lack of a signature in the $e^-p$ channel 
even with the low accumulated luminosity, supports the idea that the 
leptoquark is of the $F=0$ type. We also showed that 
future HERA measurements in all $e^{\pm}_{L,R}p$ channels will allow a 
determination of the leptoquark's quantum numbers if sufficient luminosity and
polarized beams
become available. Thirdly, we analyzed the sensitivity of the Drell-Yan 
process at the Tevatron, dijet production at LEP II as well as precision 
electroweak measurements to the existence of 
leptoquarks. In all cases we found little sensitivity to leptoquarks with 
masses near 200 GeV with values of $\tilde \lambda$ 
near 0.1. Fourthly, we found that 
the single leptoquark production process at the Tevatron Main Injector may 
provide an independent determination of its Yukawa coupling, provided that 
sufficient integrated luminosity is obtainable, while pair production at 
the NLC allows one to directly determine all the leptoquark quantum numbers.  
Lastly, we saw that leptoquarks can be embedded into a GUT structure both with 
and, surprisingly, without SUSY. Successful SUSY unification combined with 
the requirements of asymptotic freedom forced us to 
look beyond standard $SU(5)$ to the flipped $SU(5)\times U(1)_X$ model where 
$R_{2R}$ can be embedded. Without SUSY, a model with 2 Higgs doublets and 
a pair of $\widetilde R_{2L}$ leptoquarks was found to unify near $10^{15}$ GeV 
and led to reasonable values for both $\alpha_s(M_Z)$ and proton lifetime.
We hope that this excess of events at HERA is confirmed by enlarged data
samples.

\noindent{\Large\bf Acknowledgements}

We appreciate comments from 
S. Eno(D0), S. Hagopian(D0), G. Landsberg(D0), R. Harris(CDF) and 
K. Maeshima(CDF) regarding the current Tevatron 
constraints on the mass of the first generation scalar leptoquark. 
We thank P. Schlepper(H1), D. Williams(ZEUS), 
W. Smith(ZEUS), D. Krakauer(ZEUS), S. Ritz(ZEUS), and M. Derrick(ZEUS) 
for discussions of the HERA data.  
We also thank
H Dreiner, M. Peskin, D. Pierce, J. Wells, M. Worah for informative
discussions.

\section{Appendix:  Relevant Formulae}

Here we collect the relevant formulae for the processes described in the text.

\vspace{3mm}
\noindent $\bullet$ Vector Leptoquark Single Production at Hadron Colliders
\vspace{2mm}

The $\kappa$ dependent parton-level cross 
section for the single production of vector leptoquarks at hadron colliders is 
given by ($z=\cos \theta$) 
\begin{equation}
{d\sigma\over dz} = {\lamt^2\pi\alpha\alpha_s\over 96\shat}\beta\left[ 
v_1+v_2+v_3(v_4+v_5)
\right] \,,
\end{equation}
with
\begin{eqnarray}
v_1 & = & 16\left({x_2\over x_1}+{x_4\over m^2}\right) \,, \nonumber\\
v_2 & = & {8\over (2x_4+m^2)^2}\left[ a(2x_5x_6-x_4m^2)+2bx_1x_2
+2c(x_1x_6+x_2x_5-x_3x_4)-2dx_4m^2\right] \,,\nonumber\\
v_3 & = & {2\over x_1(2x_4+m^2)} \,,\\
v_4 & = & 32x_4x_5+16x_2x_5-16x_1x_4-32x_1x_6+16x_3x_4-16\kappa x_1x_2 
\,,\nonumber\\
v_5 & = & {2x_1\over m^2}\left[ -2x_5x_6+(1+\kappa)(x_2x_5+x_3x_4+x_1x_6)
-x_4m^2\right] \,,\nonumber
\end{eqnarray}
and the definitions,
\begin{eqnarray}
a & = -1+{\mbox{$2\kappa x_3$}\over \mbox{$m^2$}}\,, & b 
= -2-2\kappa+\kappa^2\,, \\
c & = 2+4\kappa+{\mbox{$(1+\kappa+\kappa^2)x_3$}\over \mbox{$m^2$}}\,, & 
d=4-{4x_3\over m^2}
-\left[ (1+\kappa)x_3\over m^2\right]^2 \,,\nonumber
\end{eqnarray}
with
\begin{eqnarray}
\begin{array}{ll}
x_1  = {1\over 2}\shat \,, & x_2=-{1\over 2}\uhat \,,\\
x_3  = {1\over 2}(m^2-\that)\,, & x_4=-{1\over 2}\that \,,\\
x_5  = {1\over 2}(m^2-\uhat)\,, & x_6={1\over 2}(\shat-m^2) \,,\\
\hat t  = -{1\over 2}(\shat-m^2)(1+z)\,, & \hat u = -{1\over 2}(\shat-m^2)
(1-z) \,.
\end{array}
\end{eqnarray}

\vspace{3mm}
\noindent $\bullet$ SM and Leptoquark Contributions to Drell-Yan Production
\vspace{2mm}

The SM contribution
to the even and odd kinematic functions are
\begin{eqnarray}
F^+_{q(SM)} & = & 2(1+z^2) \sum_{i,j} (v^i_ev^j_e+a^i_ea^j_e)(v^i_qv^j_q+
a^i_qa^j_q)P_{ij} \,,\\
F^-_{q(SM)} & = & 4z \sum_{i,j} (v^i_ea^j_e+a^i_ev^j_e)(v^i_qa^j_q+a^i_qv^j_q)
P_{ij} \nonumber \,,
\end{eqnarray}
where the sum extends over the $\gamma$ and $Z$.  Here we define
\begin{equation}
P_{ij} = {\shat^2[(\shat-M_i^2)(\shat-M_j^2)+M_i\Gamma_iM_j\Gamma_j]
\over [(\shat-M_i^2)^2+M_i^2\Gamma_i^2][(\shat-M_j^2)^2+M_j^2\Gamma_j^2]} \,, 
\end{equation}
with $M(\Gamma)$ being the masses(widths) of the gauge bosons,
and the couplings are normalized as
\begin{eqnarray}
\begin{array}{ll}
 v_f^\gamma =  Q_f \,,  & a_f^\gamma  =  0\,,\\
 v_f^Z =   \left[ {\mbox{$\sqrt 2 G_FM_Z^2$}\over \mbox{$4\pi\alpha$}}
\right]^{1/2}(T_3 -2Q_fx_w)\,, & a_f^Z  = 
\left[ {\mbox{$\sqrt 2 G_FM_Z^2$}\over \mbox{$4\pi\alpha$}}
\right]^{1/2}T_3 \,. 
\end{array} 
\end{eqnarray}
The scalar leptoquark contributions to these kinematic functions are
\begin{eqnarray}
F^+_{q(LQ)} & = & - \sum_{i=\gamma,Z}\left[ (v_e^i+a_e^i)
(v_q^i+a_q^i)\tilde\lambda^2_{L,q}+(v_e^i-a_e^i)(v_q^i-a_q^i)
\tilde\lambda^2_{R,q}\right] P_i \left[{\that^2\over \shat(\that-m^2)}+
{\uhat^2\over \shat(\uhat-m^2)}\right] \nonumber \\
& & + {1\over 4}\left[ \tilde\lambda^4_{L,q}+\tilde\lambda^4_{R,q}\right]
\left[ {\that^2\over(\that-m^2)^2} + {\uhat^2\over(\uhat-m^2)^2} \right] \,, \\
F^-_{q(LQ)} & = & \mp \sum_{i=\gamma,Z}\left[ (v_e^i+a_e^i)
(v_q^i+a_q^i)\tilde\lambda^2_{L,q}+(v_e^i-a_e^i)(v_q^i-a_q^i)
\tilde\lambda^2_{R,q}\right] P_i
\left[ {\uhat^2\over \shat(\uhat-m^2)}-{\that^2\over \shat(\that-m^2)} \right]
\nonumber\\
& & \pm {1\over 4}\left[ \tilde\lambda^4_{L,q}+\tilde\lambda^4_{R,q}\right]
\left[ {\uhat^2\over(\uhat-m^2)^2} - {\that^2\over(\that-m^2)^2} \right] 
\,.\nonumber
\end{eqnarray}
Here,
\begin{equation}
P_i  =  {s(s-M_i)\over (s-M_i^2)^2+M_i^2\Gamma_i^2} \,, 
\end{equation}
and the top sign in the anti-symmetric function
corresponds to the $F=0$, $S$-type leptoquark exchange while
the bottom sign is for the $F=-2$, $R$-type leptoquark.  Here, $q$ represents
either u- or d-quarks, depending on the coupling structure of the exchanged
leptoquark.

\vspace{3mm}
\noindent $\bullet$ SM and Leptoquark Contributions to $\epem\to q\bar q$
\vspace{2mm}

\begin{eqnarray}
{d\sigma\over dz}& = & {3\pi\alpha^2\over 2s}\left\{ \sum_{ij}\left[
(v^i_ev^j_e+a^i_ea^j_e)(v^i_qv^j_q+a^i_qa^j_q)(1+z^2)+
2(v^i_ea^j_e+v^j_ea^i_e)(v^i_qa^j_q+v^j_qa^i_q)z\right] \right. \nonumber\\
& & - {C\over 2}\sum_{i=\gamma,Z}\left[ (v_e^i+a_e^i)
(v_q^i+a_q^i)\tilde\lambda^2_{L,q}+(v_e^i-a_e^i)(v_q^i-a_q^i)
\tilde\lambda^2_{R,q}\right] P_i \\
& & \quad\quad\quad\quad \times \left\{ {t^2\over s(t-m^2)}+{u^2\over s(u-m^2)}
\pm \left[ {u^2\over s(u-m^2)}-{t^2\over s(t-m^2)} \right]\right\} \nonumber\\
& & \left. + {C\over 8}\left[ \tilde\lambda^4_{L,q}+\tilde\lambda^4_{R,q}\right]
\left\{ {t^2\over(t-m^2)^2} + {u^2\over(u-m^2)^2} \pm \left[ {u^2\over(u-m^2)^2}
- {t^2\over(t-m^2)^2} \right]\right\} \right\} \nonumber
\end{eqnarray}
with the coupling normalizations and $P_{ij}\,, P_i$ as defined in Eqs.
(27,28,30),
and the $\pm$ sign is for
\begin{equation}
\pm\quad {\rm for}\quad \left\{ \begin{array}{cc}
 S, & (F=-2)\\
 R, & (F=0)\,.
\end{array} \right.
\end{equation}

\newpage

%
\def\MPL #1 #2 #3 {Mod. Phys. Lett. {\bf#1},\ #2 (#3)}
\def\NPB #1 #2 #3 {Nucl. Phys. {\bf#1},\ #2 (#3)}
\def\PLB #1 #2 #3 {Phys. Lett. {\bf#1},\ #2 (#3)}
\def\PR #1 #2 #3 {Phys. Rep. {\bf#1},\ #2 (#3)}
\def\PRD #1 #2 #3 {Phys. Rev. {\bf#1},\ #2 (#3)}
\def\PRL #1 #2 #3 {Phys. Rev. Lett. {\bf#1},\ #2 (#3)}
\def\RMP #1 #2 #3 {Rev. Mod. Phys. {\bf#1},\ #2 (#3)}
\def\ZPC #1 #2 #3 {Z. Phys. {\bf#1},\ #2 (#3)}
\def\IJMP #1 #2 #3 {Int. J. Mod. Phys. {\bf#1},\ #2 (#3)}

\end{document}